\documentclass[10pt,journal,compsoc]{IEEEtran}
\usepackage{amsmath,amsfonts}
\usepackage{array}
\usepackage{makecell}
\usepackage{textcomp}
\usepackage{stfloats}
\usepackage{url}
\usepackage{verbatim}
\usepackage{graphicx}
\usepackage[caption=false,font=footnotesize,labelfont=rm,textfont=rm]{subfig}
\usepackage{float}
\usepackage{amsthm,amsmath,amssymb}
\usepackage{mathrsfs}
\usepackage{epstopdf}
\usepackage[linesnumbered,algoruled,boxed,lined]{algorithm2e}
\usepackage[noend]{algpseudocode}
\usepackage[section]{placeins}
\allowdisplaybreaks[4]

\usepackage{color}
\usepackage{bm}

\graphicspath{ {images/} }
\def\BibTeX{{\rm B\kern-.05em{\sc i\kern-.025em b}\kern-.08em
    T\kern-.1667em\lower.7ex\hbox{E}\kern-.125emX}}

\begin{document}

    \title{Joint Optimization of UAV-Carried IRS for Urban Low Altitude mmWave Communications with Deep Reinforcement Learning}
    

    \author{Wenwen~Xie,
    Geng Sun,~\IEEEmembership{Senior Member,~IEEE,}       
        Bei Liu,
    	Jiahui~Li, 
        Jiacheng~Wang,\\
        Hongyang Du,
        Dusit Niyato,~\IEEEmembership{Fellow,~IEEE,}
        Dong In Kim,~\IEEEmembership{Fellow,~IEEE}

    \thanks{This work is supported in part by the National Natural Science Foundation of China (62272194, 62172186, 62471200), in part by the Science and Technology Development Plan Project of Jilin Province (20230201087GX), in part by the Postdoctoral Fellowship Program of CPSF (GZC20240592), and in part by China Postdoctoral Science Foundation General Fund (2024M761123) (Corresponding authors: Geng Sun and Jiahui Li.)

    \par Wenwen Xie, Bei Liu, and Jiahui Li are with the College of Computer Science and Technology, Jilin University, Changchun 130012, China (e-mails: xieww22@mails.jlu.edu.cn; liubei0630@foxmail.com; lijiahui@jlu.edu.cn).
    
    \par Geng Sun is with the College of Computer Science and Technology, Jilin University, Changchun 130012, China, and with Key Laboratory of Symbolic Computation and Knowledge Engineering of Ministry of Education, Jilin University, Changchun 130012, China; he is also affiliated with the College of Computing and Data Science, Nanyang Technological University, Singapore 639798 (e-mail: sungeng@jlu.edu.cn).
    
    \par Jiacheng Wang and Dusit Niyato are with the College of Computing and Data Science, Nanyang Technological University, Singapore (e-mails: jiacheng.wang@ntu.edu.sg; dniyato@ntu.edu.sg).

    \par Hongyang Du is with the Department of Electrical and Electronic Engineering, The University of Hong Kong, Hong Kong 999077, China~(email: duhy@hku.hk).

    \par Dong In Kim is with the Department of Electrical and Computer Engineering, Sungkyunkwan University, Suwon 16419, South Korea. (e-mail:dongin@skku.edu).}

    \thanks{Part of this paper appeared in IEEE ICC 2024~\cite{Sun2024}.}
      
      }


\IEEEtitleabstractindextext{
    \begin{abstract}
    \par Emerging technologies in sixth generation (6G) of wireless communications, such as terahertz communication and ultra-massive multiple-input multiple-output, present promising prospects. Despite the high data rate potential of millimeter wave communications, millimeter wave (mmWave) communications in urban low altitude economy (LAE) environments are constrained by challenges such as signal attenuation and multipath interference. Specially, in urban environments, mmWave communication experiences significant attenuation due to buildings, owing to its short wavelength, which necessitates developing innovative approaches to improve the robustness of such communications in LAE networking. In this paper, we explore the use of an unmanned aerial vehicle (UAV)-carried intelligent reflecting surface (IRS) to support low altitude mmWave communication. Specifically, we consider a typical urban low altitude communication scenario where a UAV-carried IRS establishes a line-of-sight (LoS) channel between the mobile users and a source user (SU) despite the presence of obstacles. Subsequently, we formulate an optimization problem aimed at maximizing the transmission rates and minimizing the energy consumption of the UAV by jointly optimizing phase shifts of the IRS and UAV trajectory. Given the non-convex nature of the problem and its high dynamics, we propose a deep reinforcement learning-based approach incorporating neural episodic control, long short-term memory, and an IRS phase shift control method to enhance the stability and accelerate the convergence. Simulation results show that the proposed algorithm effectively resolves the problem and surpasses other benchmark algorithms in various performances.
    \end{abstract}
    
}
\maketitle

    \section{Introduction}
    \label{sec:Introduction}
    
    \par \IEEEPARstart{T}{he} promising technologies of the sixth generation (6G) communications are expected to initiate numerous new research endeavors~\cite{2023wang6Gtutorial}. Specifically, several pivotal technologies have captured significant attention, including terahertz communication~\cite{Akyildiz2022THz}, ultra-massive multiple-input multiple-output~\cite{ning2023um}, unmanned aerial vehicle (UAV)-assisted wireless communications~\cite{zhang2024TMC}, and space-air-ground integrated networks~\cite{wang2023ntn}. These technologies are capable of achieving high data rates~\cite{Wang2023TVT}, offering substantial bandwidth, and reducing interference, which can further promote great prosperity for low altitude economy (LAE) networking. However, millimeter wave communications, integral to these technologies, confront significant challenges, such as the restricted communication ranges and severe multipath effects, particularly in urban environments with extensive obstacles. Therefore, it is crucial to reconstruct the channel conditions to mitigate the effects of these obstacles for improving the robustness of such low altitude mmWave communications.

    \par Intelligent reflecting surface (IRS) is gaining recognition as an emerging technology, and it has the potential to address the aforementioned issues~\cite{Wu2021tutorial}. Specifically, an IRS consists of numerous tiny programmable reflecting elements, which can adjust their phase to alter the propagation direction and improve channel quality, thereby optimizing wireless signal transmission without the need for extra transmitters. Extensive research has been conducted to efficiently incorporate IRS into wireless systems for various purposes, such as increasing the communication throughput~\cite{li2022throughput}, optimizing the energy efficiency~\cite{su2022energy}, and enhancing the system security~\cite{sun2023TMC}. However, the performance of these systems is often constrained by the path between the source user (SU) and the users, as well as the coverage range of the SU. Once the users move out of the IRS coverage area or behind temporary obstacles, this can lead to significant signal attenuation.

    \par To address the issues above, we consider employing a UAV to carry the IRS, thereby achieving movable and flexible network assistance. Such AIRS systems have several appealing advantages over terrestrial IRS systems. For instance, with its elevated position, the IRS has a higher probability of establishing a robust line-of-sight (LoS) connection with nodes on the ground. Moreover, the IRS can achieve panoramic or full-angle reflection, which enhances the passive beamforming gain. Furthermore, utilizing the high mobility of the UAV, the AIRS system can greatly broaden its coverage area.

    \par Nevertheless, designing this AIRS system presents several challenges. First, the mobility of the terrestrial users introduces uncertainty and dynamics into the system. Second, the IRS must be parameterized by the reflection unit, which is designed through a complex optimization process. Finally, the UAV makes decisions sequentially over time, and the impact of the actions at one moment on subsequent actions must be considered to achieve the optimum over an extended period.
    
    \par As such, this paper considers a UAV-carried IRS mmWave communication system in urban scenarios and seeks to investigate an online high-performance algorithm to deal with the dynamics and uncertainty within the system. The main contributions of this work are summarized as follows:

    \begin{itemize}
    \item \textit{UAV-Carried IRS MmWave Communication System:} We consider a UAV-carried IRS system to assist low altitude mmWave communications. In such cases, obstacles block the direct link between the SU and users, so a UAV equipped with an IRS is used to create a virtual LoS link, enhancing the channel quality. Such scenarios are common and practical in urban environments.
    
    \item \textit{Joint Optimization Problem Formulation of UAV Trajectory and IRS Beamforming:} In the considered system, we formulate a optimization problem to jointly optimize the trajectory of the UAV and the phase shifts of the IRS, thereby maximizing the transmission rate and minimizing the UAV energy consumption. Considering that the optimization problem is non-convex, and the changing UAV trajectory and time-varying channels introduce dynamic nature, the optimization problem becomes challenging to solve.
    
    \item \textit{Enhanced Deep Reinforcement Learning-based Approach:} We propose a DRL-based approach, namely, enhanced proximal policy optimization (EPPO), to solve the formulated optimization problem. First, a module of neural episodic control is introduced into EPPO to accelerate the learning process. Second, mogrifier long short term memory (LSTM) is employed to improve actor network for capturing these long-term dependencies between the state and action. Finally, an IRS phase shift control strategy is employed to optimize the IRS phase shifts, which actually reduces the dimension of the action space.  
    
    \item \textit{Performance Evaluation:} A series of simulation experiments are performed to evaluate the effectiveness of the proposed EPPO algorithm. The simulation results show that the EPPO algorithm surpasses other DRL algorithms in transmission rate and energy efficiency across different scenarios, indicating that the proposed EPPO algorithm is robust and can effectively address the formulated optimization problem.
        
 
    \end{itemize}

    \par The rest of the structure of this paper is organized as follows. Section~\ref{sec:Related Work} reviews some key related works. Section~\ref{sec:System Model} presents the models and problem formulation. Section~\ref{sec:Algorithm} introduces the proposed EPPO algorithm. Simulation results are presented in Section~\ref{sec:Simulation Results And Analysis}. Finally, the paper is concluded in Section~\ref{sec:Conclusion}. 
    
%
%
\section{Related Work}
\label{sec:Related Work}

    \par In our work, we aim to use a UAV-carried IRS to aid LAE networking by optimizing the UAV trajectory and IRS parameters. In the following, we will primarily present some key related works to illustrate the novelty of our research. 
    
    \subsection{IRS and UAV assisted Communications}
    
    \par IRS has garnered widespread attention, and the related research indicates that the use of IRS can assist communication systems in mitigating path loss within the propagation environment and enhancing the quality of wireless links. For example, the authors in~\cite{lian2024TCOM} proposed a novel IRS-assisted mmWave channel model that accounts for the impact of IRS rotation angles and the effective aperture of reflecting units. The authors in~\cite{zhang2024IOTJ} presented a model for channel state information acquisition in IRS-assisted mmWave massive MIMO systems with a mixed-ADC architecture. As illustrated in~\cite{Chang2022JSAC}, the authors explored an approach for managing sensing, communication, and control in cellular-connected UAV networks by leveraging mmWave/THz communications. They introduced a novel state-to-noise-ratio definition to address the challenge, which can optimize beam alignment for data transmission while maintaining UAV control performance. Moreover, the authors in~\cite{Bai2023TWC} designed a cooperative channel model for 6G multi-UAV communications, which incorporates 3D dynamic trajectories and self-rotation to enhance the accuracy of non-stationary massive MIMO mmWave channel predictions. However, the works above did not consider the mobility of user equipment and existence of the obstacles, nor the integration of UAV-IRS-assisted mmWave communications.
    
    \par Recently, AIRS or UAV-carried IRS has demonstrated advantages that terrestrial IRS does not possess, such as a wide coverage angle, strong LoS links, and high mobility. Specifically, an innovative scheme for IRS-UAV-assisted wireless communication was designed in~\cite{Su2022TCOM}, where an IRS attached to a UAV to is utilized to improve the quality of the air-to-ground channel. In~\cite{lu2021TWC}, the authors introduced a novel three-dimensional wireless system architecture enabled by AIRS. Unlike conventional terrestrial IRS, AIRS provides enhanced deployment flexibility and a wider signal reflection coverage. Likewise, the authors in~\cite{wei2023TCOM} designed a secure communication scheme supported by an AIRS, and developed an iterative algorithm to maximize he worst-case sum secrecy rate by optimizing the IRS phase shifts, hovering location of the UAV, and the beamforming matrix of the access point (AP). In addition, the authors in~\cite{sun2024IOTJ} utilized multiple AIRSs to enhance the timeliness and rate of the data in IoT networks. Specifically, they aimed to minimize the average AoI and maximize the data rate by jointly designing the deployment locations of AIRSs,  phase shift matrix of the AIRSs, and transmission scheduling of IoT devices. However, these studies have not considered the dynamics and continuity of the environment, and they have not utilized online algorithms to tackle these challenges either.

    \subsection{Optimizations of IRS Parameters and UAV Trajectory}
    
    \par \par Recent studies have shown that the optimization of IRS parameters and UAV trajectory can further enhance the signal quality and improve the energy efficiency of UAV-assisted IRS systems. For example, in~\cite{Wei2021}, a frequency division multiple access (OFDMA)-based IRS-assisted UAV orthogonal communication system was proposed, which enhances the overall communication rate by leveraging the beamforming gain of the IRS and the high mobility of the UAV. Moreover, a secure UAV network assisted by IRS in the presence of an eavesdropper is investigated in~\cite{pang2022TCOM}, where the average secrecy rate was maximized by optimizing the trajectory of the UAV, the beamforming matrix of the transmitter, and the phase shift of the IRS, thereby enhancing the security of wireless communications. Likewise, the authors in~\cite{WangJinming2023TVT} designed a novel UAV-driven IRS-BackCom for multiple users, where the UAV serves as a power source, and the IRS reflects and modulates electromagnetic waves to convey data to the users. In addition, the authors in~\cite{Wang2024TVT} presented an innovative approach to IRS-assisted UAV uplink transmission, focusing on enhancing energy efficiency through the optimization of UAV trajectory, user scheduling, and IRS phase shifts. However, these studies did not account for the simultaneous optimization of both the UAV trajectory and IRS beamforming when the UAV is equipped with an IRS.

    \subsection{Optimization Methods for IRS and UAV Communications} 
    
    \par DRL has been increasingly recognized for its real-time adaptability in wireless communications, presenting promising opportunities to dynamically optimize communication system performance and resource utilization efficiency. For example, an IRS-assisted cognitive UAV network is studied in~\cite{deng2024IOTJ}, and the authors formulated a non-convex optimization problem to maximize the data throughput through adjusting the sensing duration, IRS passive beamforming, and the three-dimensional location of the UAV. They addressed the non-convex optimization problem by leveraging bisection search, closed-form phase shifts, and convex approximation, resulting in an efficient algorithm for a sub-optimal solution. Similarly, the authors in~\cite{zhai2022TWC} proposed a game-theoretic model for an IRS-assisted wireless communication network with wireless power delivery and multicast transmissions. Specifically, the authors utilized two schemes based on Stackelberg game theory to enhance the utilities of both the PS and transmitter, and they developed an alternating optimization algorithm with low complexity to determine energy pricing and wireless energy transmission time, achieving better performance than schemes without IRS support. Moreover, a UAV-assisted mobile edge computing system is investigated in~\cite{song2023TMC}, and the corresponding optimization objectives are minimizing the task completion delay and UAV energy consumption while maximizing the quantity  of the completed tasks. To solve this, the authors applied an evolutionary multi-objective RL algorithm which leverages population evolution to maintain multiple promising RL policies simultaneously, thereby improving the performance of the algorithm.
    \par Moreover, The authors in~\cite{joint2020} employed the deep Q-Network and deep deterministic policy gradient (DDPG) algorithm to maximize the energy efficiency of the UAV in a traditional IRS-assisted UAV communication system. The authors in~\cite{Huang2020risa} investigated the downlink MISO system with multiple users, and developed an an efficient algorithm using DRL that simultaneously adjusts the transmit beamforming and phase shifts, focusing on improving the throughput performance of the system. In contrast, an IRS-assisted aerial-to-ground uplink NOMA cellular network is studied in~\cite{Zhao2022raga}. With the objective of maximizing the overall data rate, they utilized a distributed and robust DRL algorithm to jointly optimize the UAV trajectory, IRS passive beamforming, and power control. Furthermore, the authors in~\cite{Liu2022ppob} investigated a THz-frequency integrated sensing and communication system. Specifically, they formulated an optimization problem with ergodic constraints and used distributed DRL to enhance both capacity and performance. Moreover, Furthermore, the authors in~\cite{ji2023tacd} proposed an innovative framework for cellular networks with cache-enabled multi-UAVs and formulated a partially observable stochastic game to reduce the overall content acquisition latency using DRL. However, these studies only considered real-time state information and did not account for long-term dependencies.
    
    \subsection{Summary} 
    \par In summary, the distinctions between our work and previous research can be outlined as follows. First, certain previous studies overlooked potential obstacles, which may reduce the likelihood of establishing strong LoS links during low altitude mmWave communications. Second, many works did not integrate the mobility of the UAV with the IRS. Third, the optimization methods of these studies did not adequately manage the trade-offs between immediate and long-term benefits. Different from the existing works, we consider a UAV-carried IRS mmWave communication system in urban low altitude communication scenarios and seek to investigate an online high-performance algorithm capable of addressing the dynamics and uncertainty of the system.

%
%
\section{System Model}
\label{sec:System Model}

    \begin{table}[t]
    \caption{Main Notations.}
    	\label{tab:Main Notation}
    	\renewcommand{\arraystretch}{1.3}
    	\begin{center}
    		
    		\begin{tabular}{|c|c|} 
    			\hline
    			\textbf{Notation}                              & \textbf{Definition}                                            \\
                    \hline
    			\makecell[c]{$a^x_t,a^y_t,a^z_t$}              & \makecell[c]{ Flying distances of UAV in time slot $t$ }       \\
                    \hline
    			\makecell[c]{$d_0$}                            & \makecell[c]{ Drag ratio }                                     \\
                    \hline
    			\makecell[c]{$d^{SI}_t$}                       & \makecell[c]{ Distance between SU and IRS in time slot $t$ }   \\
    			\hline
    			\makecell[c]{$d^{IE}_t$}                       & \makecell[c]{ Distance between IRS and user in time slot $t$ } \\
                    \hline
    			\makecell[c]{$D^{max}$}        & \makecell[c]{ Maximal flying distances of UAV }                \\
                    \hline
    			\makecell[c]{$g_t$}                            & \makecell[c]{ Channel gain of SU-IRS link in time slot $t$ }   \\
                    \hline
    			\makecell[c]{$G$}                              & \makecell[c]{ Rotor disc area }                                \\
                    \hline
    			\makecell[c]{$h_t$}                            & \makecell[c]{ Channel gain of IRS-user link in time slot $t$ } \\
                    \hline
    			\makecell[c]{$M_r,M_c$}                        & \makecell[c]{The number of reflecting elements of IRS          \\ in each row and column} \\
                    \hline
    			\makecell[c]{$n$}                            & \makecell[c]{ Total number of user } \\
                    \hline
    			\makecell[c]{$P,\sigma^2,B$}                   & \makecell[c]{ Transmission power, noise power, bandwidth }     \\
                    \hline
    			\makecell[c]{$R_t$}                            & \makecell[c]{ Data rate of SU-IRS-user link in time slot $t$ } \\
                    \hline
    			\makecell[c]{$s$}                              & \makecell[c]{ The rotor solidity }                             \\
    			\hline
    			\makecell[c]{$t,T,\mathcal{T}$}                & \makecell[c]{The index, the number, and the set                \\ of time slots} \\ 
    			\hline
    			\makecell[c]{$t_d$}                            & \makecell[c]{ Time duration of time slot }                     \\
                    \hline
    			\makecell[c]{$U_{tip}$}                        & \makecell[c]{ Tip speed of the rotor blade }                   \\
    			\hline
    			\makecell[c]{$v_0$}                            & \makecell[c]{ Mean rotor induced velocity in hover }           \\
    			\hline
    			\makecell[c]{$x^{SU},y^{SU},z^{SU}$}           & \makecell[c]{Coordinate of SU}                                 \\
    			\hline
    			\makecell[c]{$x^{IRS}_t, y^{IRS}_t,z^{IRS}_t$} & \makecell[c]{ Coordinate of UAV-carried IRS in time slot $t$ }             \\
    			\hline
    			\makecell[c]{$x^{UE}_{u,t},y^{UE}_{u,t},z^{UE}_{u,t}$}     & \makecell[c]{ Coordinate of the $u$-th user in time slot $t$ }            \\
    			\hline
    			\makecell[c]{$X^{min},X^{max}$}                & \makecell[c]{ Border of target area on the x-axis }            \\
    			\hline
    			\makecell[c]{$Y^{min},Y^{max}$}                & \makecell[c]{ Border of target area on the y-axis }            \\
    			\hline
    			\makecell[c]{$Z^{min},Z^{max}$}                & \makecell[c]{ The minimal, maximal of                          \\ flying altitude of UAV }\\
    			\hline
    			\makecell[c]{$\alpha$}                         & \makecell[c]{ Air density }                                    \\
    			\hline
    			\makecell[c]{$\Theta_t$}                       & \makecell[c]{ Phase shift matrix of IRS in time slot $t$ }     \\
    			\hline
    		\end{tabular}
    	\end{center}
    \end{table}

\par In this section, we first present the overview of the considered UAV-carried IRS mmWave communication system in urban low altitude communication scenarios. Then, we detail the UAV energy consumption and communication models to derive the decision variables of transmission performance and energy efficiency of the system. The main notations used in this paper are summarized in Table \ref{tab:Main Notation}.

\subsection{System Overview}
\label{subsec:System Overview}

    \par
\begin{figure}[t]
        \centerline{\includegraphics[width=3.5in]{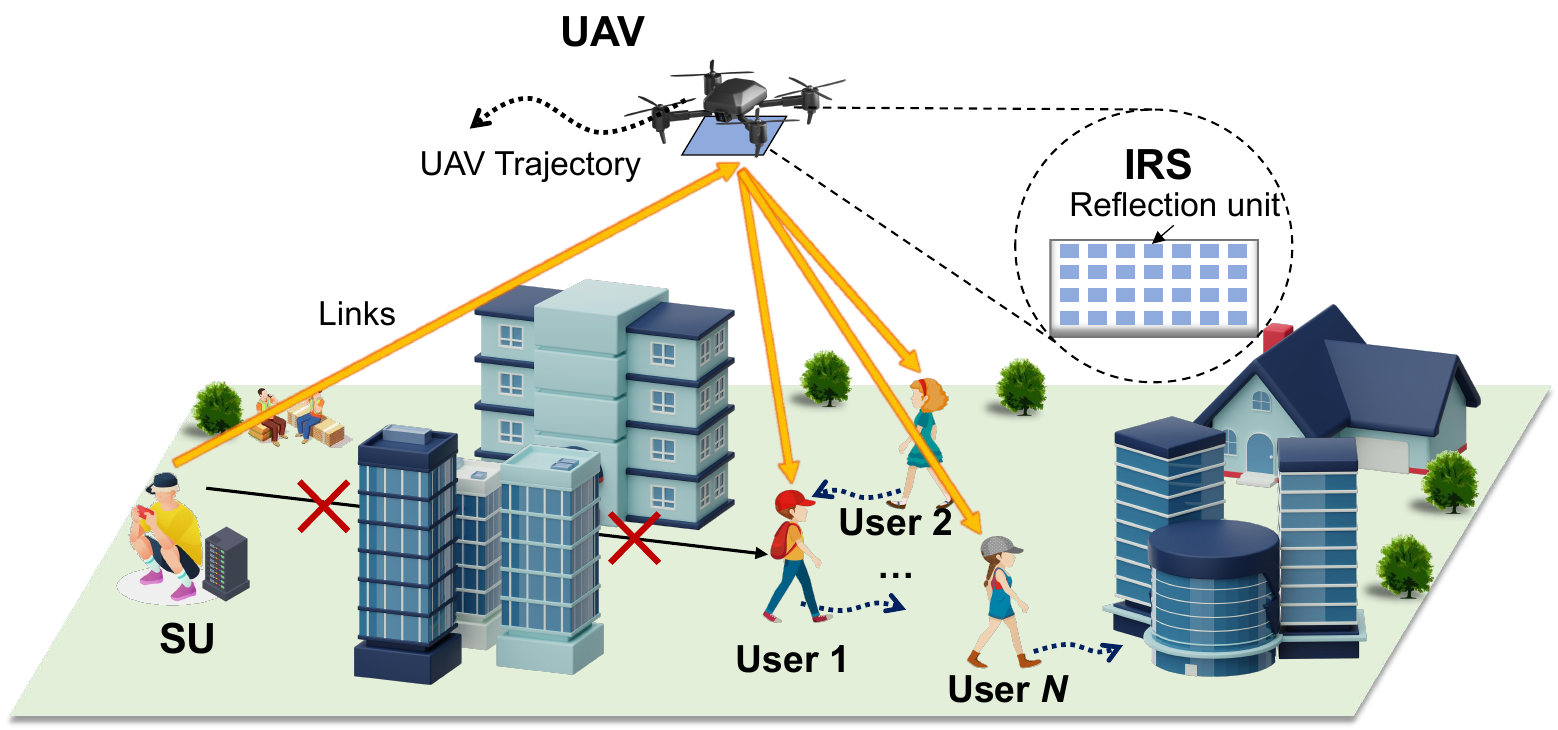}}
        \caption{UAV-carried IRS-aided mmWave communication system in urban low altitude communication scenarios.}
        \label{fig.system_model}
\end{figure}

 As shown in Fig.~\ref{fig.system_model}, we consider a mmWave communication system consisting of an SU, a UAV-carried IRS with $M = M_r \times M_c$ reflecting elements, and mobile users. Specifically, the SU and mobile users are both in an urban low altitude communication scenario, which means that their direct links are easily blocked by obstacles. In this case, a UAV-carried IRS is dispatched to establish the links from the SU to each user. As such, the signal will first reach the reflection element of the IRS, and then be reflected by the IRS to the mobile user. Since large high-rise buildings cause severe path loss and high attenuation, we considered that the LoS links between the SU and the users are unavailable if there are any obstacles in the middle. Moreover, akin to~\cite{Du2021, Ye2022}, we consider that the SU and mobile users do not employ MIMO, because the highly dynamic nature of the urban low-altitude economic scenario may compromise robustness with its introduction. Note that this model can be easily extended to the MIMO systems by embedding the existing MIMO methods~\cite{zheng2021double,zhang2024IOTJ}.

    
    \par We consider a discrete-time system evolving over time slots $\mathcal{T} \triangleq \{1,2,\ldots,T\}$, where the length of the time period is equal to $t_d$ seconds. For the sake of simplicity, we consider the user to be served with a time-division-multiple-access (TDMA) mode. Without loss of generality, we consider a Cartesian coordinate system, where the locations of the SU, UAV-carried IRS and the $u$-th user at time slot $t$ are represented as $\mathbf{c}^{SU} \triangleq [x^{SU},y^{SU},z^{SU}]$, $\mathbf{c}^{IRS}_t \triangleq [x^{IRS}_t, y^{IRS}_t,z^{IRS}_t]$, $\mathbf{c}^{UE}_{u,t} \triangleq[x^{UE}_{u,t},y^{UE}_{u,t},z^{UE}_{u,t}]$, respectively.

    \par Due to the mobility of the user, the UAV-carried IRS needs to change its position to achieve better communication performance. In the following, we will introduce the SU-IRS-user communication models, and then present the UAV mobile and energy cost models.

    \subsection{Communication Model}
    \label{subsec:Communication Model}
    
    \par We consider the Rician fading channels for the communication links between the SU and IRS as well as between the IRS and the user, because IRS can transform Rayleigh/fast fading into Rician/slow fading to achieve ultra-high reliability~\cite{Wu2021tutorial}. Since the location of the UAV-carried IRS and the user varies with time, the path loss between the SU and UAV-carried IRS and the path loss between the UAV-carried IRS and $u$-th user in time slot $t$ are given by $\Psi^{SI}_t = \beta(D) + 10n\log_{10}(d^{SI}_t/D)$ and $\Psi^{IE}_{u,t} = \beta(D) + 10n\log_{10}(d^{IE}_{u,t}/D)$~\cite{Jiang2021}, respectively, where $d_0$ is the reference distance, and $\beta(D)$ is the reference path loss at a reference distance $D$. Moreover, $n$ is the path loss exponent, which determines the rate of signal strength decay with distance. 

    \par Following this, we use $g_t \in \mathbb{C}^{M \times 1}$ to denote the downlink channel vector from the SU to the UAV-carried IRS, and $h_{u,t} \in \mathbb{C}^{M \times 1}$ to denote the downlink channel vector from the UAV-carried IRS to the user in time slot $t$, which are respectively given by
    \begin{equation}
    \label{eq:the downlink channel vector from the BS to the IRS}
    \begin{split}
        g_t \quad= \Psi^{SI}_t(\sqrt{\frac{k}{1+k}}\widetilde{g}^{LoS}_t+\sqrt{\frac{1}{1+k}}\widetilde{g}^{NLoS}_t),
    \end{split}
    \end{equation}
    \begin{equation}
    \label{eq:the downlink channel vector from the IRS to the user}
    \begin{split}
        h_{u,t} = \Psi^{IE}_{u,t}(\sqrt{\frac{k}{1+k}}\widetilde{h}^{LoS}_{u,t}+\sqrt{\frac{1}{1+k}}\widetilde{h}^{NLoS}_{u,t}),
    \end{split}
    \end{equation}
    \noindent where $k$ is the Rician factor, $\Psi^{IE}_{u,t}$ and $\Psi^{SI}_t$ are the path loss from the UAV-carried IRS to the user and the path loss from the SU to the UAV-carried IRS, respectively. Moreover, the non-LoS (NLoS) part of channel $\widetilde{h}^{NLoS}_{u,t}$ and $\widetilde{g}^{NLoS}_t$ are generally modeled as \textit{i.i.d.} standard Gaussian distributions. The LoS part of channel $\widetilde{h}^{LoS}_{u,t}$ and $\widetilde{g}^{LoS}_t$ are related to the locations of the SU, the UAV-carried IRS, and the $u$-th user. Furthermore, we use $\varphi^{SI}_t$ and $\psi^{SI}_t$ to denote the azimuth and elevation angles of arrival (AoA) from the SU to the UAV-carried IRS, $\varphi^{IE}_{u,t}$ and $\psi^{IE}_{u,t}$ to denote the azimuth and elevation angles of departure (AoD). Then, the LoS channel $\widetilde{h}^{LoS}_{u,t}$ and $\widetilde{g}^{LoS}_t$ can be expressed as follows~\cite{10070838}:
    \begin{equation}
    \label{eq:the LoS component from the SU to the IRS}
    \begin{split}
        \widetilde{g}^{LoS}_t = [1,\ldots,e^{\frac{2\pi jl}{\lambda}\{ m_c\sin{(\varphi^{SI}_t)}\cos{(\psi^{SI}_t)} + m_r\sin{(\psi^{SI}_t)} \}},\ldots, \\ e^{\frac{2\pi j}{\lambda}\{ (M_c-1)\sin{(\varphi^{SI}_t)}\cos{(\psi^{SI}_t)} + (M_r-1)\sin{(\psi^{SI}_t)} \}}],
    \end{split}
    \end{equation}
    \begin{equation}
    \label{eq:the LoS component from the IRS to the user}
    \begin{split}
        \widetilde{h}^{LoS}_{u,t} = [1,\ldots,e^{\frac{2\pi jl}{\lambda}\{ m_c\sin{(\varphi^{IE}_{u,t})}\cos{(\psi^{IE}_{u,t})} + m_r\sin{(\psi^{IE}_{u,t})} \}},\ldots, \\ e^{\frac{2\pi jl}{\lambda}\{ (M_c-1)\sin{(\varphi^{IE}_{u,t})}\cos{(\psi^{IE}_{u,t})} + (M_r-1)\sin{(\psi^{IE}_{u,t})} \}}],
    \end{split}
    \end{equation}
    \noindent where $l$ is the distance between two adjacent reflecting elements of the UAV-carried IRS, $\lambda$ is the carrier wavelength, $m_r$ is the row index of reflecting elements and $m_c$ is the column index of reflecting elements. Moreover, $M_r$ is the total number of reflecting elements in a row and $M_c$ is the total number of reflecting elements in a column.  

    
    
    
    %
    \par Finally, we let $\Theta_t = \text{diag} \{ e^{j\omega_{1,t}},e^{j\omega_{2,t}},\ldots,e^{j\omega_{M,t}} \}$ be the reflection coefficients at the UAV-carried IRS, where $\omega_{i,t} \in [-\pi,\pi)$ is the phase shift of the $i$-th element. Then, the achievable data rate $R_{u,t}$ at the $u$-th user in the time slot $t$ can be expressed as follows:
    \begin{equation}
    \label{eq:data rate}
    \begin{split}
        R_{u,t} = B\log_2{(1 + \frac{Pg_t^T\Theta_th_{u,t}}{B\sigma^2})},
    \end{split}
    \end{equation}
    \noindent where $P$ and $\sigma^2$ are the transmission and noise power, respectively. Moreover, $B$ is the bandwidth.

  \subsection{UAV Mobile and Energy Cost Models}
    
    \par In each time slot, the UAV moves with a flying action $a_t \triangleq [a^x_t,a^y_t,a^z_t]$. Thus, the coordinate of the UAV in time slot $t$ is calculated by $\mathbf{c}^{UAV}_t = \mathbf{c}^{UAV}_{t-1} + a_t$. The distance between the SU and UAV-carried IRS in time slot $t$ is expressed as $d^{SI}_t = \Vert \mathbf{c}^{SU} - \mathbf{c}^{IRS}_t \Vert$. Similarly, the distance between the UAV-carried IRS and $u$-th user is $d^{IE}_{u,t} = \Vert \mathbf{c}^{IRS}_t - \mathbf{c}^{UE}_{u,t} \Vert$.

    \par Then, we introduce the following energy cost model for the mobile UAV. Specifically, the main energy consumption of the UAV is propulsion energy consumption since the UAV is used to carry the IRS to reflect signals and does not participate in the communication process. Thus, for a rotary UAV flying in a three-dimensional (3D) space, the propulsion energy consumption in time slot $t$ is represented as follows~\cite{joint2020}:
    \begin{equation}
    \label{eq:the energy consumption of UAV}
    \begin{split}
        E_t =& (P_B \big(1+\frac{3({v^h_t})^2}{U^2_{tip}}\big)+ P_I \big(\sqrt{1+\frac{({v^h_t})^4}{4v^4_0}}-\frac{({v^h_t})^2}{2v^2_0} \big)^{\frac{1}{2}}
        \\&+\frac{1}{2}d_0{\alpha}sG({v^h_t})^3 + mg{v^v_t}) t_d,
    \end{split}
    \end{equation}
    \noindent where the constants $P_B$ and $P_I$ represent the blade profile power and induced power in hovering status, respectively. The tip speed of the rotor blade is denoted by $U_{tip}$, and $v_0$ refers to the mean rotor-induced velocity during hover. The fuselage drag ratio and rotor solidity are represented by $d_0$ and $s$, respectively, while $\alpha$ and $G$ denote air density and rotor disc area, respectively. Additionally, $m$ stands for the total mass of the UAV and IRS, and $g$ is the gravitational acceleration. The horizontal and vertical velocities of the UAV in time slot $t$ are denoted by $v^h_t = \sqrt{(a^x_t)^2 + (a^y_t)^2} / t_d$ and $v^v_t = |a^z_t| / t_d$, and $t_d$ is the time duration.
    

    \subsection{Problem Formulation}
    \label{subsec:Problem Formulation}

    \par In this work, the system focuses on two primary goals which are optimizing the transmission rate and minimizing the energy consumption of the UAVs. However, in multi-user scenarios, a UAV-carried IRS may serve only a few users, which can result in high transmission rates for some and low transmission rates for others. This imbalance may degrade the overall system performance and negatively impact user experience. Therefore, we seek to define a new objective function that can jointly optimize the transmission performance, energy efficiency, and fairness as the optimization objective. 

    \par Specifically, we first introduce Jain's fairness index~\cite{jain1984quantitative}, a metric used to evaluate the fairness of resource allocation, to ensure that all users achieve reasonable transmission rates. By balancing the data rates among users, this metric can maximize the overall system transmission rate while minimizing the rate differences among users. In particular, the fairness coefficient $\xi$ is defined as follows:
    \begin{equation} 
    \label{eq
    's Fairness Index} 
    \begin{aligned} \xi = \frac{(\sum\nolimits^n_{i=1}R_{i,t})^2}{n\sum\nolimits^n_{i=1}R_{i,t}^2} \in \left[\frac{1}{n}, 1\right], \end{aligned} 
    \end{equation} 
    \noindent where $n$ represents the total number of users. Then, we combine Jain's fairness index with transmission rate and energy consumption, and thus define the fairness rate energy consumption ratio as the optimization objective. The fairness rate energy consumption ratio at the $t$th time slot is given by 
    \begin{equation}
    \label{eq: f_objective}
        F_t = \frac{\sum^n_{i=1}\xi R_{i,t}}{E_t}.
    \end{equation}
    
    \par Based on this, the phase adjustments of the IRS and the flight path of the UAV should be jointly controlled to improve the transmission rate and fairness coefficient. Moreover, aimed at optimizing the energy consumption of the UAV, we need to plan the trajectory of the UAV to minimize the overall energy usage. As such, the following decision variables need to be jointly determined: \textit{(i)} $\boldsymbol{\Theta} = \{\Theta_t | t \in \mathcal{T}\}$, a diagonal matrix representing the reflection coefficients of the UAV-carried IRS for different time periods; \textit{(ii)} $\boldsymbol{A} = \{[a^x_t,a^y_t,a^z_t] | t \in \mathcal{T}\}$, a matrix representing the control parameters of the UAV, which denotes its spatial displacement at different time intervals.

    \par Following this, the corresponding joint optimization problem is formulated as follows:
    \begin{subequations}
    \label{eq:problem formulation}
    \begin{align}
        \max \limits_{\boldsymbol{\Theta},\boldsymbol{A}}\quad &{\sum^T_{t=1} F_t} \\ 
        \text{s.t.} \quad  &C1: 0 \leq d_t^{IRS} \leq D^{max} \label{subeq:problem formulation 1} \\ 
        &C2: X^{min} \leq x^{IRS}_t \leq X^{max} \label{subeq:problem formulation 2}\\ 
        &C3: Y^{min} \leq y^{IRS}_t \leq Y^{max} \label{subeq:problem formulation 3}\\ 
        &C4: Z^{min} \leq z^{IRS}_t \leq Z^{max} \label{subeq:problem formulation 4}\\ 
        &C5: -\pi \leq \omega_{i,t} < \pi,\  i = 1,\ldots,M \label{subeq:problem formulation 5}
    \end{align}
    \end{subequations}
    \noindent where Eq.~\eqref{subeq:problem formulation 1} represents the constraint on the flying distance of the UAV-carried IRS per second. Moreover, Eqs.~\eqref{subeq:problem formulation 2}, ~\eqref{subeq:problem formulation 3}, and~\eqref{subeq:problem formulation 4} define the permissible flight areas for the UAV-carried IRS, with any flight outside these areas considered a boundary violation. In addition, Eq.~\eqref{subeq:problem formulation 5} restricts the range of phase shifts available to the UAV-carried IRS. Note that this problem is non-convex, and the non-convexity arises from the reflection coefficients $\boldsymbol{\Theta}$. Specifically, since the LoS links need to be reconstructed based on the current position, the value range for $\boldsymbol{\Theta}$ is constrained by the current location, resulting in the selection space for $\boldsymbol{\Theta}$ non-convex. Moreover, even if we only consider the IRS optimizing in case other variables are known, the simplified problem can be reduced as a non-convex quadratically constrained quadratic program (QCOP), which is proven to be an NP-hard problem~\cite{Wu2021tutorial}. Thus, the formulated problem is also NP-hard and cannot be solved in polynomial time. 
    
    

    \section{DRL-based Approach}
    \label{sec:Algorithm}
    
    \par In this section, we propose a DRL-based approach to solve our joint optimization problem. To this end, we first show the motivations for using DRL and reformulate the problem as a markov decision process (MDP). Then, we introduce the proposed EPPO algorithm with several improvements.
    
    
    \subsection{Motivations for Using DRL and MDP Formulation}
    \label{subsec:Motivations}

    \par Our formulated problem is dynamic and uncertain since it involves a highly unpredictable environment, where factors such as obstacles, UAV mobility, and user position can change rapidly. Moreover, the UAV-carried IRS system also exhibits high requirements for real-time response. Thus, the commonly used static optimization methods such as convex or non-convex optimization are not suitable for this problem~\cite{wang2023stn, Li2023TMC, Li2022SECON}. In this case, DRL is able to swiftly adapt the dynamic and uncertain environment and offer robust solutions. Thus, we seek to adopt the DRL method to solve our optimization problem.  

    \par To this end, we first reformulate the optimization problem shown in Eq.~(\ref{eq:problem formulation}) as an MDP. Mathematically, an MDP is a tuple ($\mathcal{S}, \mathcal{A}, \mathcal{P}, R, \gamma$) which are state space, action space, state transition probability, reward function, and discount factor, respectively. Among them, state, action, and reward are the most important components which are detailed as follows.


    \subsubsection{State Space}
    \label{subsubsec:State Space}
    
    \par The state space is designed to encompass critical spatial factors that impact system performance. Specifically, the coordinates of the UAV and the $u$-th user are contained since these parameters affect the channel conditions. As such, the state $s_t$ is given by $s_t = \{\mathbf{c}^{UAV}_t, \mathbf{c}^{UE}_{u,t}\}$.

    \subsubsection{Action Space}
    \label{subsubsec:Action Space}
    
    \par In our system, the UAV can adjust the position to achieve better channel conditions. Moreover, the UAV-carried IRS also can turn its parameters in terms of phase shifts and reflection angles, which can be computed by the coordinates of the SU, the UAV, and the user. As such, the action of agent $a_t$ is represented as $a_t = \{a^x_t,a^y_t,a^z_t \}$, where $a^x_t,\ a^y_t$ and $a^z_t$ are the flight distance of UAV along the $x$, $y$, and $z$ axes.

    \subsubsection{Reward Function}
    \label{subsubsec:Reward Function}
    
    \par A well-designed reward function contributes to problem-solving, which is critical. As such, the reward function incorporates both our optimization objective and the associated constraints, as outlined in Eq.~(\ref{eq:reward function}). Specifically, this function motivates the UAV to achieve better transmission rates through high LoS probabilities while minimizing the energy consumption within the flyable areas, \textit{i.e.}, 
    \begin{equation}
    \label{eq:reward function}
    \begin{split}
        \left\{
        \begin{aligned}
            r_t &= \xi R_{u,t}/E_t - p_{o}\ && \text{if\ LoS\ link} \\
            r_t &= 0\ && \text{if\ NLoS\ link}. 
        \end{aligned}
        \right.
    \end{split}
    \end{equation}
    \noindent where $p_o$ is the out-of-bounds penalty, and $\xi$ is Jain’s Fairness Index. In what follows, we aim to propose a DRL algorithm to handle this reformulated MDP. 
    
    

    \subsection{PPO Algorithm}
    \label{subsec:PPO Algorithm}
    \par In this work, we aim to consider PPO as the solving framework. Specifically, PPO~\cite{ppo2017} is a state-of-the-art reinforcement learning algorithm based on policy. This type of policy-based algorithm will generate a policy network $\pi_\theta$ to make decisions in the abovementioned MDP. Thus, the objective of PPO is to improve the policy parameters for achieving high state values, \textit{i.e.}, 
    \begin{equation}
    \label{eq:the policy gradient objective function}
    \begin{split}
         J(\theta) = \mathbb{E}_{\tau\sim\pi_\theta}\left[\sum\nolimits^T_{t=1}r_t\right].
    \end{split}
    \end{equation}
    
    \par Policy gradient methods face a limitation in sample efficiency. This inefficiency stems from the need to sample multiple complete trajectories $\tau$, which is computationally expensive, particularly in the scenarios involving high-dimensional state spaces or continuous action spaces. To mitigate this, the actor-critic method is introduced, thereby improving sample efficiency. Specifically, the actor-critic method incorporates a value function, referred to as the critic, which assesses the effectiveness of the current action. This assessment is then used to guide updates to the actor network in a more efficient manner. The actor-critic method aims to maximize the objective of the actor while minimizing the loss function of the critic. A hyperparameter $\beta_b$ is used to balance the actor and critic network during the learning process. \textit{i.e.},
    \begin{equation}
    \label{eq:the a-c objective function}
    \begin{split}
        J(\theta,V) = J_{actor}(\theta) - \beta_bJ_{critic}(V),
    \end{split}
    \end{equation}
    
    \noindent where $J_{actor}(\theta) = \mathbb{E}_{\tau\sim\pi_{\theta}} \left[\sum\nolimits^T_{t=1} \nabla_\theta \log \pi_\theta (a_t | s_t) A^{\pi_\theta}(s_t, a_t) \right]$. In Eq.~(\ref{eq:the a-c objective function}), $\nabla_\theta \log \pi_\theta (a_t | s_t)$ represents the policy gradient with respect to the parameter $\theta$, and $A^{\pi_\theta}(s_t, a_t)$ denotes the advantage function used to estimate the benefit of taking a specific action relative to the average action. This advantage function can be either the temporal difference error or an estimate derived from the value function. Moreover, the critic network seeks to minimize the square of the temporal difference error, where $\gamma$ is the discount factor, and $V(s_t)$ represents the state value estimate by the critic, $J_{critic}(V) = \mathbb{E}_\tau \left[ \sum\nolimits^T_{t=1} \frac{1}{2} (r_t + \gamma V(s_{t+1}) - V(s_t))^2 \right]$.

    \par Trust region policy optimization (TRPO) is an enhancement of the policy gradient algorithm within the actor-critic framework. Specifically, TRPO introduces a trust region to ensure that each parameter update does not cause excessive policy changes, thereby enhancing the algorithmic stability. This approach reduces the risk of introducing overly large policy variations during updates, which can lead to learning instability. The objective function of TRPO comprises two components: one is for the objective of the actor and another one is for the constraint term that restricts policy changes. The overall objective function of TRPO is expressed as follows:
    \begin{equation}
    \label{eq:the TRPO objective function}
    \begin{aligned}
        \max \quad &J(\theta) = \mathbb{E}_{\tau\sim\pi_{\theta}} \left[\sum\nolimits^T_{t=1} \nabla_\theta \log \pi_\theta (a_t | s_t) A^{\pi_\theta}(s_t, a_t) \right], \\
        \text{s.t.}\quad  &\mathbb{E}_{\tau\sim\pi_{\theta_{old}}} \left[ D(\pi_{\theta_{old}} (\cdot | s) || \pi_\theta(\cdot|s)) \right] \leq \delta ,
    \end{aligned}
    \end{equation}
    \noindent where $\delta$ is a pre-defined threshold and $D(\cdot || \cdot)$ represents Kullback-Leibler divergence.
    \par PPO imposes a limit on the proportion between the probabilities of the new and previous policies, ensuring that policy updates are within a safe and bounded range. To this end, PPO improves the policy using a surrogate objective that constrains policy updates. Let $A_{\pi_\theta}$ be the advantage function, then the constraint can be expressed as follows:  
    \begin{equation}
    \label{eq:the clipped surrogate objective function}
    \begin{split}
        L^{clip}(\theta) = \mathbb{E}_t[\min( \rho_t(\theta)A_{\pi_\theta}(s_t,a_t),\rho^{clip}_t(\theta)A_{\pi_\theta}(s_t,a_t) )],
    \end{split}
    \end{equation}

    \noindent where $\theta$ represents the policy parameters, $\rho_t(\theta)$ is the ratio of new and old policy probabilities, and $\rho^{clip}_t(\theta) = clip(\rho_t(\theta),1-\epsilon,1+\epsilon)$ is a clipped value of $\rho_t(\theta)$, where $\epsilon$ is a hyperparameter that controls the size of the policy update. As such, PPO combines this surrogate objective with multiple epochs of data to iteratively update the policy while avoiding large policy deviations, resulting in stable learning. 
    
    \par However, the instantaneous variation of the reward function may be large in our MDP. Thus, it is hard for conventional PPO to catch the policy in the short term, which causes PPO to exhibit slow convergence. In this case, we aim to improve the PPO alforithm so that it better aligns with our MDP in the following.

    \subsection{Ehanced PPO}
    \label{subsec:EPPO}

    \par An improved version of the PPO algorithm, referred to as EPPO, is introduced in this subsection. Specifically, the large and continuously changing dynamic state space poses a convergence challenge for the algorithm, as it struggles to generalize and adapt to such a complex environment. Moreover, our considered system evolves over time, where the actions taken at each time step have a lasting impact on subsequent states and actions. Therefore, the agent is required to understand the relationships and influences that span across extended periods for accurate predictions and decision-making. However, PPO is not well-suited to capture these long-term dependencies between states and actions. Additionally, the vast and multi-dimensional action space increases exploration difficulty and computational complexity, making it harder for the algorithm to efficiently learn an optimal policy. Thus, EPPO integrates three essential enhancements which are


    \begin{itemize}
        \item Neural episodic control with state abstraction (NECSA): NECSA is to reduce the state and action space, simplifying and accelerating the learning process.
        \item Mogrifier LSTM: Mogrifier LSTM can handle long-sequence dependencies, enabling the algorithm to learn better.
        \item IRS phase shift control strategy: IRS phase shift control strategy can reduce the dimension of the action space.
    \end{itemize}
    They are detailed as follows. 

    \subsubsection{Neural Episodic Control with State Abstraction}
    \label{subsubsec:NECSA}

    \par
\begin{figure}[tb]
    		\centerline{\includegraphics[width=3.5in]{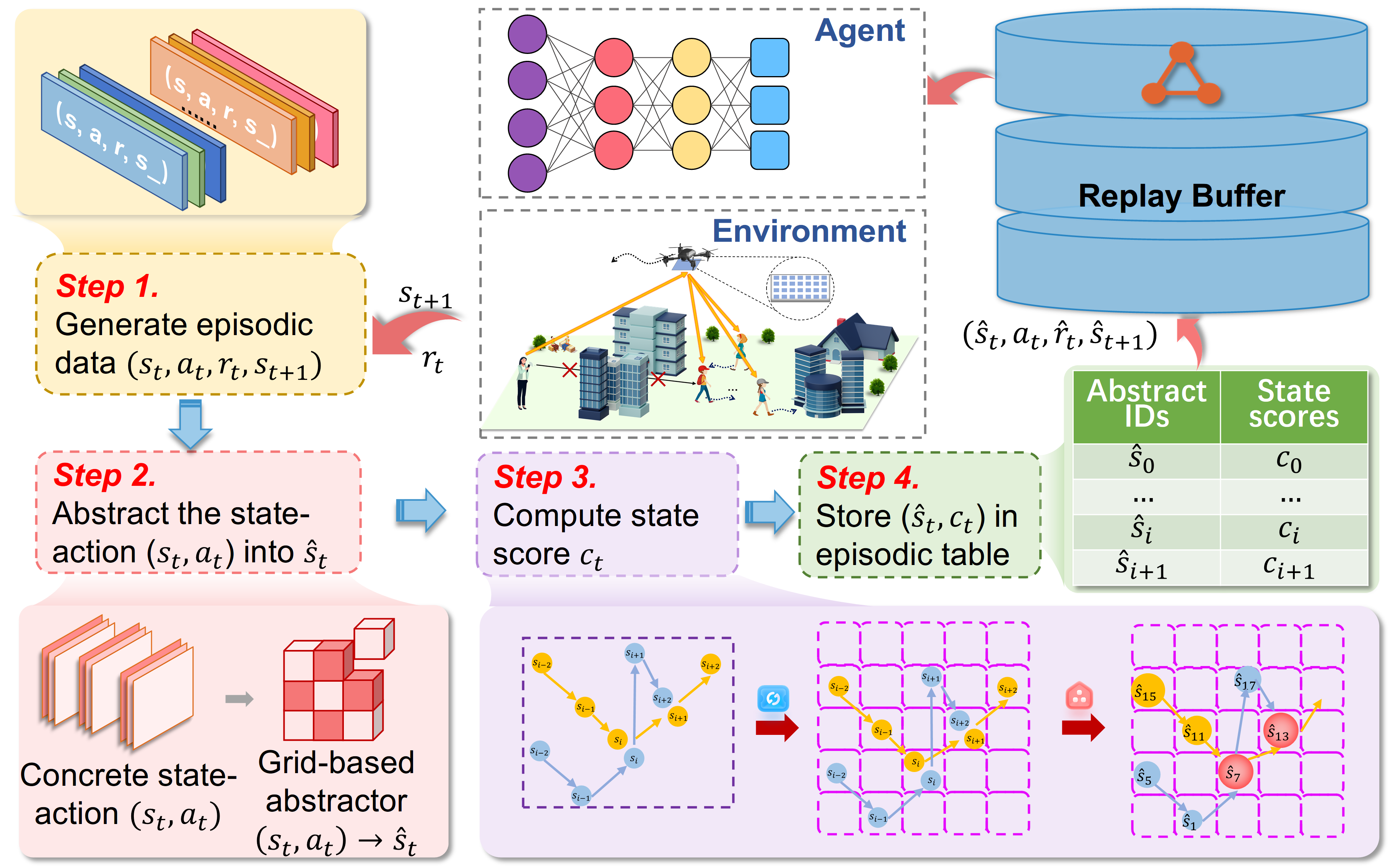}}
    		\caption{The overview of NECSA.}
    		\label{fig:NECSA overview}
    \end{figure}

 To speed up the convergence, we first introduce NECSA~\cite{li2023neural} mechanism into our EPPO algorithm. Fig.~\ref{fig:NECSA overview} illustrates that NECSA is mainly composed of three essential components that are the state abstractor, the episodic table, and the replay buffer. First, the state abstraction plays a pivotal role, transforming the continuous state space into a grid-based abstract representation. This process aids efficient learning in high-dimensional spaces. Second, the episodic table stores scores linked to abstract states, which are continuously updated as the agent interacts with the environment. Finally, the memory buffer stores past experiences, allowing the agent to compare its current state with these memories for improved learning. As an improvement of PPO algorithm, NECSA takes a transition ($s_t, a_t, r_t, s_{t+1}$) from the environment and processes it using the three mentioned modules. Subsequently, it generates a new transition ($s_t, a_t, \hat{r}_t, s_{t+1}$), where $\hat{r}_t$ represents an intrinsic reward calculated based on the episodic table and the original reward $r_t$. This reward revision enhances the learning process by providing more precise and informative feedback to the agent, ultimately improving its performance.
    
    \subsubsection{Mogrifier LSTM}
    \label{subsubsec:Mogrifier LSTM}
    \par
\begin{figure}[tb]
    		\centerline{\includegraphics[width=3.5in]{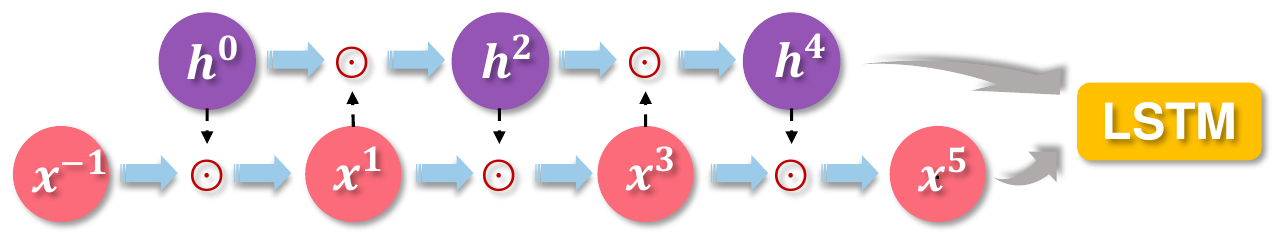}}
    		\caption{Mogrifier LSTM with 5 rounds of updates.}
    		\label{fig:Mogrifier LSTM with 5 rounds of updates}
    \end{figure}

 To achieve a higher reward, we introduce Mogrifier LSTM~\cite{Melis2020} to effectively capture long-range dependencies and contextual information of our MDP. Specifically, mogrifier LSTM is a variant of LSTM that introduces an innovative computation step before the standard LSTM operation. This process is depicted in Fig.~\ref{fig:Mogrifier LSTM with 5 rounds of updates}, where inputs $x$ and $h$ undergo interactive transformations in an alternating method. Moreover, $\odot$ represents the element-wise product. Mogrifier LSTM enhances the actor neural network of PPO. The actor neural network is designed as a deep neural network, combining a multi-layer perceptron and a mogrifier LSTM, producing action probabilities from the given observed environment state. In addition, the critic neural network remains unchanged.
    
    \subsubsection{IRS Phase Shift Control Strategy}
    \label{subsubsec:Phase Control Strategy}
    
    \par To further facilitate the convergence, A low-complexity algorithm is proposed to optimize the phase shifts of the UAV-carried IRS, aiming to maximize user data rate using the known coordinates of both the UAV and the user at time slot $t$. The optimal phase shift $\omega_{i,t}$ is denoted as~\cite{Wei2021}:
    \begin{equation}
    \label{eq:phase shift optimization}
    \begin{split}
        \omega_{i,t} = & \frac{2\pi l}{\lambda} \{ (m_c - 1)\sin{\varphi^{IE}_t}\cos{\psi^{IE}_t} + (m_r - 1)\sin{\psi^{IE}_t} \\
        &+ (m_c - 1)\sin{\varphi^{SI}_t}\cos{\psi^{SI}_t} + (m_r - 1)\sin{\psi^{SI}_t}\},
    \end{split}
    \end{equation}
    \noindent where $m_c$ and $m_r$ represent the column and row indices of the $i$th element in the UAV-carried IRS, respectively. PPO includes a process for learning the phase shifts of UAV-carried IRS, which expands the action space and slows down the convergence. In contrast, the IRS phase control strategy can effectively reduce the action space by calculating the phase shifts using Eq.~(\ref{eq:phase shift optimization}). Specifically, the original action space of $(M_r \times M_c+3)$ is reduced to 3.

    \begin{figure*}[tb]
    		\centerline{\includegraphics[width=7in]{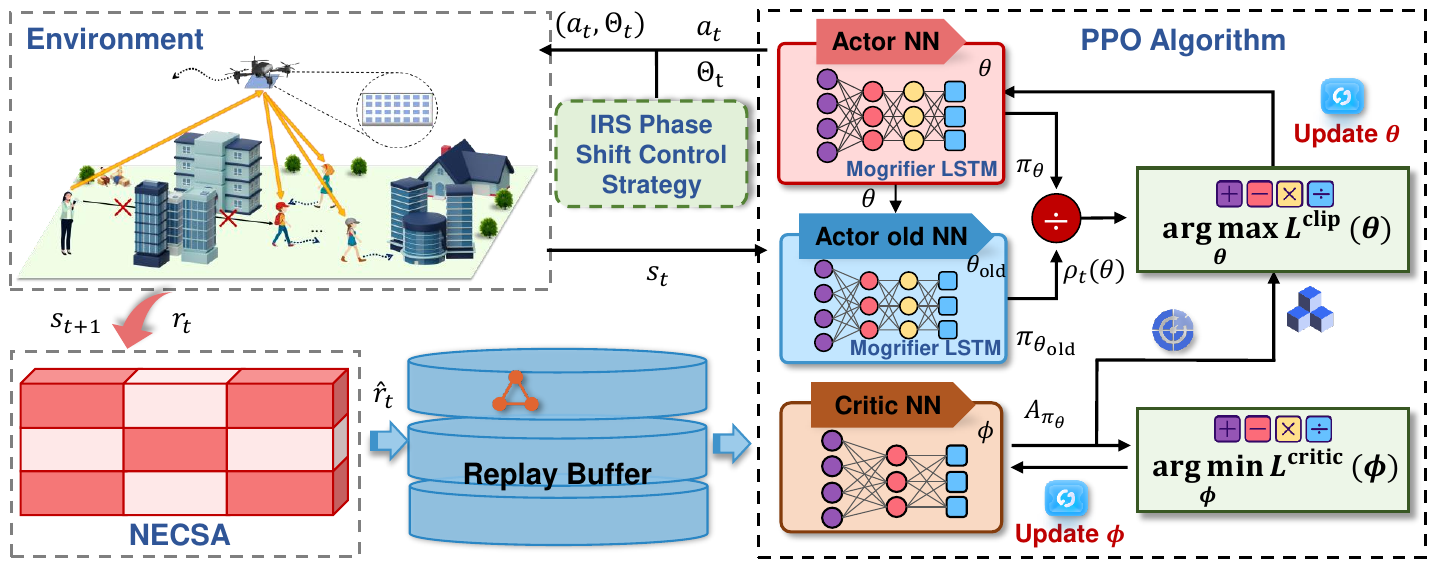}}
    		\caption{The framework of the the proposed EPPO.}
    		\label{fig:The framework of the EPPO algorithm.}
    \end{figure*}
    
    \subsubsection{Main Steps of EPPO Algorithm}
    \label{subsubsec:EPPO Algorithm}
    \begin{algorithm}[tb]
		\caption{EPPO}
		\label{Algorithm 1}
    \SetAlgoNoEnd
    \SetAlgoNoLine
        Initialize the parameter $\theta$ of actor neural network and the parameter $\phi$ of critic neural network;\\
        Initialize the parameter of actor\_old neural network $\theta_{old} \leftarrow \theta$; \\
        Initialize the episodic memory $C$ \\
        \For{ Episode = $1,\ldots,N^{eps}$} 
        {
            Reset the environment and initialize state $s_t$; \\
            \For{Time slot $t = 1,\ldots,T$}
            {
                Obtain state $s_t$; \\
                Select action $a_t = \{a^x_t, a^y_t, a^z_t \}$; \\
                Execute action $a_t$; \\
                Calculate the energy consumption of UAV $E_t$ from Eq.~(\ref{eq:the energy consumption of UAV}); \\
                Obtain the optimized phase shifts of UAV-carried IRS according to Section \ref{subsubsec:Phase Control Strategy}; \\
                Calculated $r_t$ according to Eq.~(\ref{eq:reward function}); \\
                Revised reward $\hat{r}_t = NECSA([s_t,a_t,r_t,s_{t+1}],C)$; \\
                Store transition $[s_t,a_t,\hat{r}_t,s_{t+1}]$ into experience replay buffer; \\
                \If{Size(replay buffer) == batch size}
                {
                    Copy parameter of actor neural network to actor\_old neural network $\theta_{old} \leftarrow \theta$; \\
                    \For{epoch $K_e = 1,\ldots,K$}
                    {
                        Update the parameter $\theta$ and parameter $\phi$; 
                    }
                    Empty the replay buffer;
                }
            }
        } 
	\end{algorithm}
    \par As depicted in Fig.~\ref{fig:The framework of the EPPO algorithm.}, EPPO employs two neural networks, \textit{i.e.}, an actor neural network which is responsible for policy learning, and a critic neural network that estimates the state value function and the advantage function. To ensure the stable policy updates, EPPO maintains an actor\_old network with identical parameters to calculate the clipping ratio. The current state $s_t$ is fed into the neural network to generate action $a_t$. Following the execution of this action, the environment yields a corresponding reward $r_t$, creating a transition. This transition is updated by the NECSA module and then added to the replay buffer, which is used for neural network training.

    \par The main steps of the proposed EPPO algorithm are shown in Algorithm~\ref{Algorithm 1}. Specifically, the EPPO algorithm begins by initializing the actor and critic neural network parameters, along with the episodic memory. Then, EPPO proceeds to execute episodes of interactions with the environment. Within each episode, it resets the environment and state, and for each time slot, the algorithm selects actions, optimizes UAV-carried IRS phase shifts, calculates UAV energy consumption, computes rewards, and revises rewards using the NECSA module. Next, the transitions are saved in an experience replay buffer for further use. When the number of transitions in the replay buffer reaches the predefined batch size, the algorithm updates the actor and critic neural network parameters through multiple training epochs, ensuring convergence. This process repeats until all specified episodes are completed, and the algorithm continuously refines its policy for reinforcement learning tasks.

    \begin{table*}
    \caption{Complexity Comparison of Different Algorithms.}
        \label{tab:Complexity Comparison of Different Algorithms.}
        \centering
        \begin{tabular}{ccc}
            \hline
            Algorithm & Computational complexity & Space complexity\\
            \hline
            EPPO & \makecell{$\mathcal{O}(|\theta_A| + |\phi_C| + N^{eps}T|\theta_A| + N^{eps}TN^{nec}V + K(|\theta_A| + |\phi_C|))$} & \makecell{$\mathcal{O}(|\theta_A| + |\phi_C| + D_r(3|s|+2|a|+3))$}\\
            DDPG & \makecell{$\mathcal{O}(2|\theta_A| + 2|\phi_C| + N^{eps}T|\theta_A| + N^{eps}TV + K(2|\theta_A| + 2|\phi_C|))$} & \makecell{$\mathcal{O}(2|\theta_A| + 2|\phi_C| + D_r(2|s|+|a|+2))$}\\
            TD3 & \makecell{$\mathcal{O}(2|\theta_A| + 4|\phi_C| + N^{eps}T|\theta_A| + N^{eps}TV + K(2|\theta_A| + 4|\phi_C|))$} & \makecell{$\mathcal{O}(2|\theta_A| + 4|\phi_C| + D_r(2|s|+|a|+2))$}\\
            SAC & \makecell{$\mathcal{O}(|\theta_A| + 4|\phi_C| + N^{eps}T|\theta_A| + N^{eps}TV + K(|\theta_A| + 4|\phi_C|))$} & \makecell{$\mathcal{O}(|\theta_A| + 4|\phi_C| + D_r(2|s|+|a|+2))$}\\
            PPO & \makecell{$\mathcal{O}(|\theta_A| + |\phi_C| + N^{eps}T|\theta_A| + N^{eps}TV + K(|\theta_A| + |\phi_C|))$} & \makecell{$\mathcal{O}(|\theta_A| + |\phi_C| + D_r(2|s|+|a|+2))$}\\
            \hline
        \end{tabular}
        
    \end{table*}

    \subsection{Complexity Analysis}
    \label{subsec:Complexity Analysis}

    \par In this section, we analyze the computational and space complexity of EPPO during training and execution phases.    
    \par \textbf{Training Phase: } The computational complexity of EPPO is $\mathcal{O}(|\theta_A| + |\phi_C| + N^{eps}T|\theta_A| + N^{eps}TN^{nec}V + K(|\theta_A| + |\phi_C|))$ in the training phase, which can be summarized as follows:
    \begin{itemize}
        \item \textbf{\textit{Network Initialize}}: This phase includes the initialization of network parameters. The corresponding computational complexity is given by $\mathcal{O}(|\theta_A| + |\phi_C|)$, where $|\theta_A|$ and $|\phi_C|$ represent the number of parameters in the actor and critic networks, respectively.
        \item \textbf{\textit{Action Sampling}}: In this phase, actions are generated based on the current state, with the associated complexity given by $\mathcal{O}(N^{eps}T|\theta_A|)$. Here, $N^{eps}$ represents the total number of training episodes, and $T$ refers to the length of time slot set $\mathcal{T}$
        \item \textbf{\textit{Replay Buffer Collection}}: The complexity of collecting state transitions in the replay buffer is $\mathcal{O}(N^{eps}TN^{nec}V)$, where $N^{nec}$ is the computational complexity of NECSA and $V$ indicates the complexity associated with environment interaction.
        \item \textbf{\textit{Network Update}}: The network parameters undergo $K$ updates during the updating phase. As a result, the complexity for this phase is calculated as $\mathcal{O}(K(|\theta_A| + |\phi_C|))$.
    \end{itemize}
    \par During training, the space complexity for EPPO is given by $\mathcal{O}(|\theta_A| + |\phi_C| + D_r(3|s|+2|a|+3))$, where $D_r$ denotes the size of the replay buffer, and $|s|$ and $|a|$ correspond to the dimensions of the state and action spaces, respectively. This space complexity also includes the size of the neural network parameters and the replay buffer that stores the tuples of ($s_t, a_t, r_t, s_{t+1}$). The complexity of other algorithms can be found in Table~\ref{tab:Complexity Comparison of Different Algorithms.}.
    \par \textbf{Execution Phase: } In the execution phase, the computational complexity associated with EPPO is $\mathcal{O}(|\theta_A|)$, which is primarily due to the actor network inferring actions based on the current state. Note that the space complexity in this phase is also $\mathcal{O}(|\theta_A|)$

    \subsection{System Deployment}
    \par In this subsection, a feasible deployment method is introduced to implement the system. In our considered system, the SU acts as the DRL agent to control the movement of the UAV-carried IRS by performing the proposed EPPO algorithm, where the link between SU and UAV can be viewed as wireless backhaul so that the UAV can be controlled by the SU. Moreover, the UAV acts as the IRS agent to manage the phase shift adjustment of the IRS. The system deployment is divided into two phases, \textit{i.e.}, positioning phase and transmission phase. The detailed process is as follows:
    \begin{itemize}
        \item \textit{Positioning Phase}: Since the position of the UAV in each time slot is computed and controlled by the SU, the positioning phase mainly focuses on acquiring the location of the mobile user. In this case, some existing works have addressed user positioning issues in mmWave communication scenarios~\cite{Peng2024, 9410435}. For instance, the positioning of mobile users in IRS-assisted mmWave communication systems is investigated in~\cite{9410435}, where the authors proposed a random beamforming and maximum likelihood estimation method to estimate some key parameters such as the AoA and AoD, enabling centimeter-level positioning accuracy for the mobile user. Therefore, such methods can be applied in the positioning phase to obtain the location of the mobile user.
        \item \textit{Transmission Phase:} According to the location of the mobile user obtained in the positioning phase, the SU can control the trajectory of the UAV in the next time slot by performing our proposed EPPO algorithm based on the observed environment state. Moreover, Given the UAV's limited energy and computational resources, the UAV can be installed a low-power lightweight field-programmable gate array (FPGA) to control the phase shifts of the IRS~\cite{10643685} referred to Eq.~(\ref{eq:phase shift optimization}). Note that the complexity of phase computation is extremely low and the FPGA is lightweight, making it feasible for the energy-limited and computation-limited UAV. Accordingly, once the UAV trajectory and IRS phase shifts are determined, the data transmission process is then completed.
    \end{itemize}

    \section{Simulation Results and Analysis}
    \label{sec:Simulation Results And Analysis}

        \begin{table}[t]
    \caption{Simulation Parameters.}
    \label{tab:Simulation Parameters}
    	\renewcommand{\arraystretch}{1.3}
    	\begin{center}
    		\begin{tabular}{|c|c|c|c|} 
    			\hline
    			\textbf{Parameter}       &  \textbf{Value}        &  \textbf{Parameter}       &  \textbf{Value}         \\
    			\hline
                \makecell[c]{$Z^{min}$}  &  \makecell[c]{80 m}  &  \makecell[c]{$Z^{max}$}  &  \makecell[c]{120 m}  \\ 
                \hline
                \makecell[c]{$X^{min}$}  &  \makecell[c]{0 m}   &  \makecell[c]{$X^{max}$}  &  \makecell[c]{620 m}  \\ 
                \hline
                \makecell[c]{$Y^{min}$}  &  \makecell[c]{0 m}   &  \makecell[c]{$Y^{max}$}  &  \makecell[c]{620 m}  \\ 
                \hline
                \makecell[c]{$T$}  &  \makecell[c]{300}  &  \makecell[c]{$D^{max}$}  &  \makecell[c]{30}  \\ 
                \hline
                \makecell[c]{$P_B$}  &  \makecell[c]{199.4}  &  \makecell[c]{$P_I$}  &  \makecell[c]{88.66}  \\ 
                \hline
                \makecell[c]{$m$}  &  \makecell[c]{2 kg}  &  \makecell[c]{$g$}  &  \makecell[c]{9.8 N/kg}  \\ 
                \hline
                \makecell[c]{$U_{tip}$}  &  \makecell[c]{120 m/s}  &  \makecell[c]{$v_0$}  &  \makecell[c]{4.03}  \\ 
                \hline
                \makecell[c]{$d_0$}  &  \makecell[c]{0.6}  &  \makecell[c]{$\alpha$}  &  \makecell[c]{1.225 $\rm kg/m^3$}  \\ 
                \hline
                \makecell[c]{$s$}  &  \makecell[c]{0.05}  &  \makecell[c]{$G$}  &  \makecell[c]{0.53 $\rm m^2$}  \\ 
                \hline
                \makecell[c]{$t_d$}  &  \makecell[c]{1 s}  &  \makecell[c]{$D$}  &  \makecell[c]{1 m}  \\ 
                \hline
                \makecell[c]{$\beta(D)$}  &  \makecell[c]{30 dB}  &  \makecell[c]{$n$}  &  \makecell[c]{2.2}  \\ 
                \hline
                \makecell[c]{$l$}  &  \makecell[c]{$\lambda/2$}  &  \makecell[c]{$P$}  &  \makecell[c]{15 W}  \\ 
                \hline
                \makecell[c]{$\sigma^2$}  &  \makecell[c]{-174 dBm/Hz}  &  \makecell[c]{$B$}  &  \makecell[c]{2 MHz}  \\ 
                \hline
                \makecell[c]{$N^{eps}$}  &  \makecell[c]{3000}  &  \makecell[c]{$\gamma$}  &  \makecell[c]{0.99}  \\ 
                \hline
                \makecell[c]{$p_o$}  &  \makecell[c]{0.04}  &  \makecell[c]{$\epsilon$}  &  \makecell[c]{0.02}  \\ 
                \hline
    		\end{tabular}
    	\end{center}
    \end{table}

    \par In this section, we present the simulation results and analyses. We first introduce the simulation setting and benchmarks, and then provide the simulation results. 
    
    
    \subsection{Simulation Setups}
    \label{subsec:Simulation Setups}


    \par We consider a typical dense urban low altitude communication scenario with multiple lanes and intersections. The UAV flies within the airspace above a square area. The area is divided into a $3 \times 3$ grid, with each grid cell measuring 10 m in length. Inside each grid cell, 8 buildings are randomly placed, and the gaps between the grid cells represent roads, each 10 m wide. Our coordinate system places the bottom left corner of the city as the origin, with SU coordinates at (-200, 0, 25) and the initial coordinates of the user at (305, 205, 0). In the multi-user scenario, the initial coordinates for the three users are (305, 205, 0), (305, 405, 0), and (305, 105, 0), respectively. Following this, the UAV is deployed randomly in a valid city area for each episode, while the user follows the road. Other parameters can be found in Table~\ref{tab:Simulation Parameters} and follow reference \cite{joint2020}.

    \par For comparison, we utilize deep DDPG~\cite{ddpg2016}, twin delayed deep deterministic policy gradient (TD3)~\cite{td32018}, soft actor-critic (SAC)~\cite{sac2018}, and proximal policy optimization (PPO)~\cite{ppo2017} as benchmark methods.

    \begin{itemize}
        \item DDPG: DDPG is a DRL algorithm designed for solving problems in high dimensional states and continuous action spaces. Specifically, it utilizes a deterministic policy to estimate the optimal policy while maintaining a target network to stabilize training.
        \item TD3: TD3 is an improved version of DDPG, and it introduces two Q-networks to reduce overestimation errors and employs a double delayed strategy to enhance training stability. TD3 is particularly robust in high-noise and uncertain environments, improving the accuracy of Q-value estimates.
        \item SAC: SAC is a DRL algorithm that leverages the maximum entropy principle to encourage the agent to explore the action space. Specifically, it introduces the concept of Soft Q-Value and entropy to balance exploration and exploitation, making it suitable for highly uncertain environments. SAC exhibits strong exploration capabilities and can adaptively adjust the trade-off between exploration and exploitation.
        \item PPO: PPO is an on-policy reinforcement learning algorithm, and it adopts a probabilistic policy and updates policy parameters through proximal optimization, resulting in high stability. PPO can dynamically adjust the learning rate during training to enhance robustness.
    \end{itemize}

    \subsection{Optimization Results in Single-User Scenario}
    \label{subsec:Optimization Results for single user}

     \begin{figure*}
      \centering           
      \subfloat[]   
      {
          \label{fig:Cumulative rewards training curve.}\includegraphics[width=0.325\textwidth]{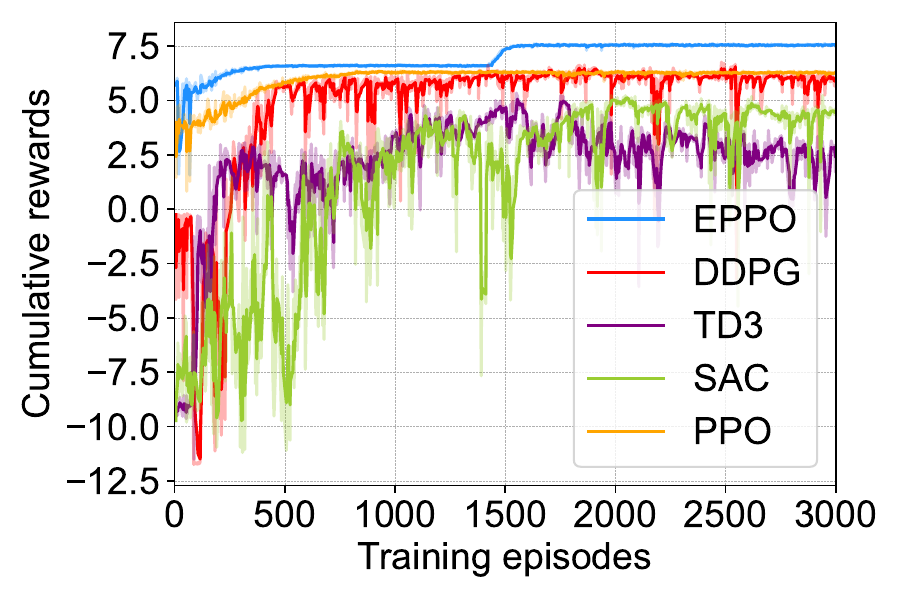}
      }
      \subfloat[]
      {
          \label{fig:Average rate training curve.}\includegraphics[width=0.325\textwidth]{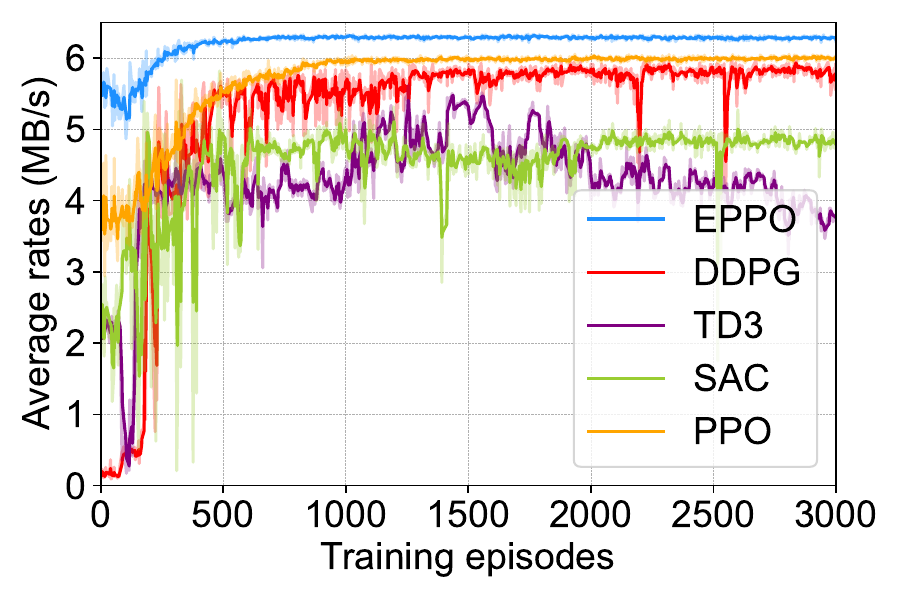}
      }
      \subfloat[]
      {
          \label{fig:Cumulative energy consumption training curve.}\includegraphics[width=0.325\textwidth]{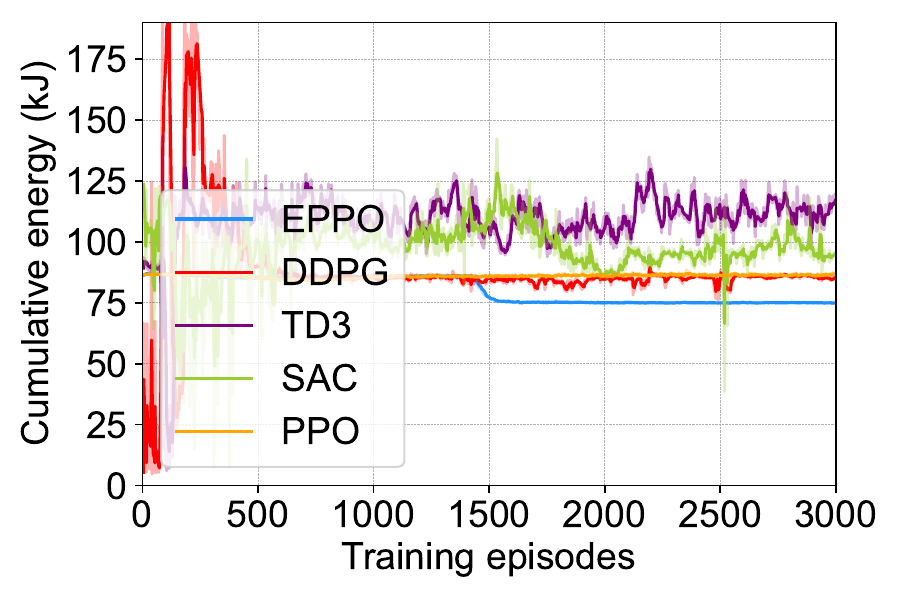}
      }
      \caption{Training results in the single-user scenario. (a) Cumulative rewards training curve. (b) Average rate training curve. (c) Cumulative energy consumption training curve.}    
      \label{fig:simulation results of EPPO}            
    \end{figure*}


    \begin{figure*}
      \centering  
      \subfloat[]   
      {
          \label{fig:Comparison of mogrifier LSTM, LSTM and GRU.}\includegraphics[width=0.24\textwidth]{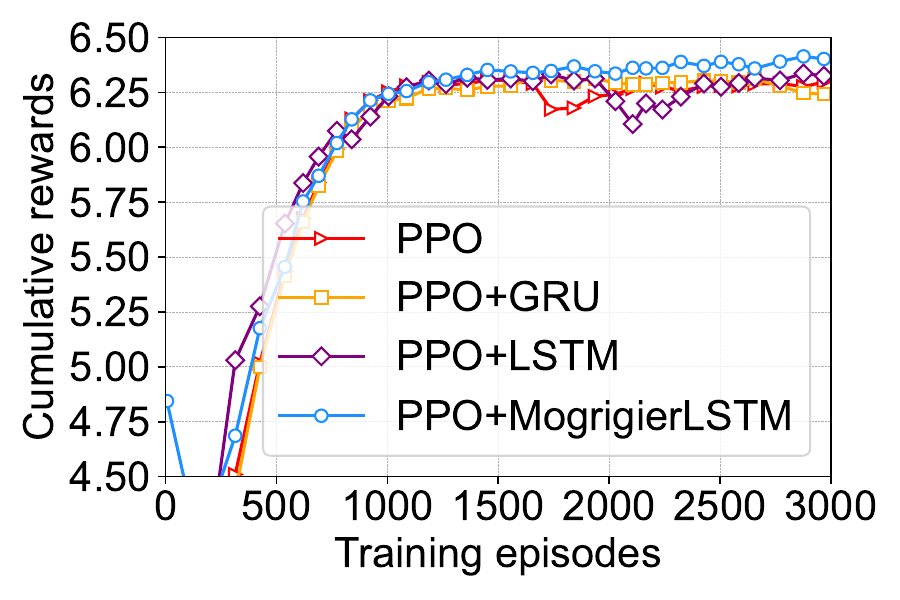}
      }
      \subfloat[]   
      {
          \label{fig:Comparison of PPO and PPO with NECSA.}\includegraphics[width=0.24\textwidth]{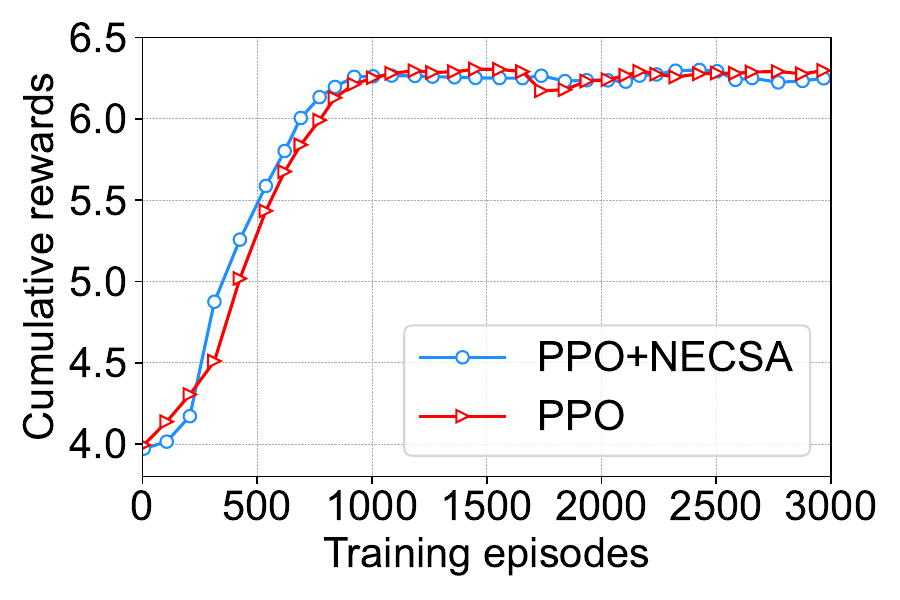}
      }
      \subfloat[]
      {
          \label{fig:Comparison of PPO and PPO with phase control.}\includegraphics[width=0.24\textwidth]{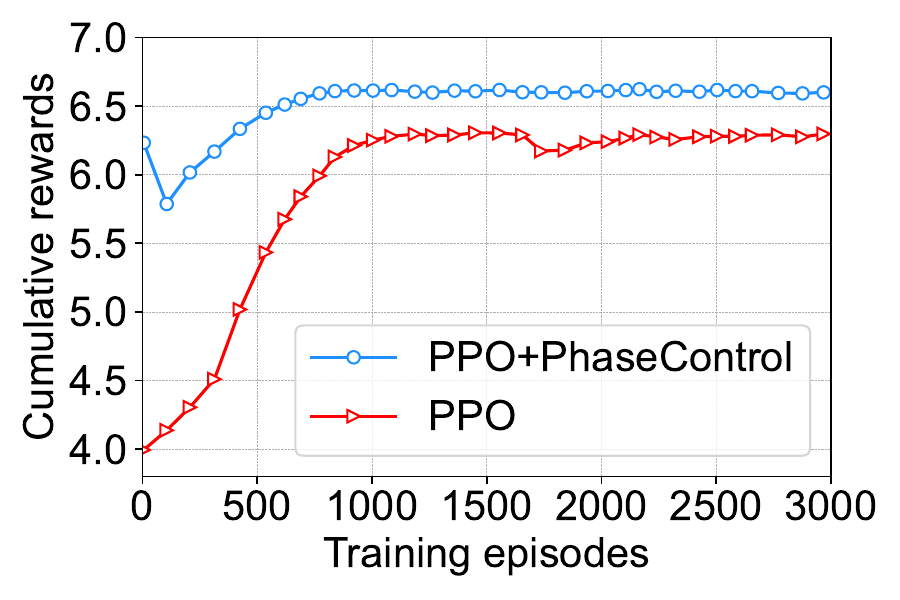}
      }
      \subfloat[]
      {
          \label{fig:PPO_PPO_PhaseControl_MogrifierLSTM.}\includegraphics[width=0.24\textwidth]{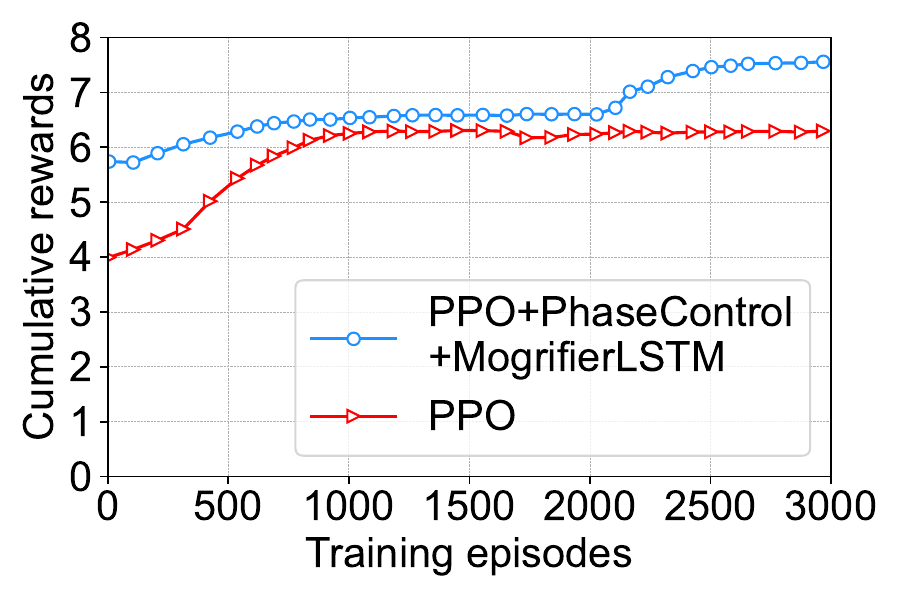}
      }
      \caption{Ablation simulation results in the single-user scenario. (a) Comparison of mogrifier LSTM, LSTM and GRU. (b) Comparison of PPO and PPO with NECSA. (c) Comparison of PPO and PPO with IRS phase shift control. (d) Comparison of PPO and PPO with IRS phase shift control and mogrifier LSTM.}    
      \label{fig:comparsion results of improvement}            
    \end{figure*}

    \begin{figure}
    		\centerline{\includegraphics[width=2.5in]{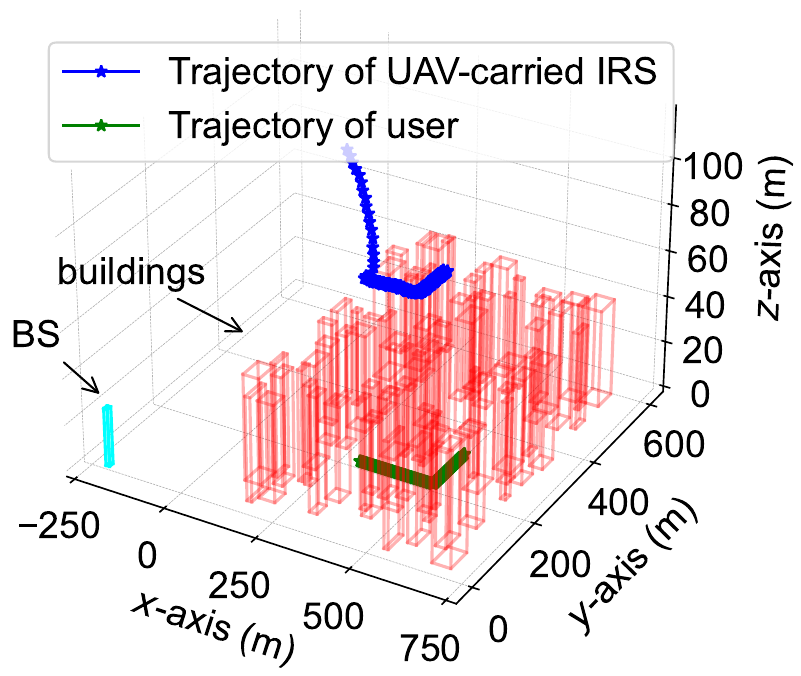}}
    		\caption{3D trajectories of the UAV for single user.}
    		\label{fig:UAV 3D Trajectory of EPPO for single user.}
    \end{figure}

    \par In this subsection, we first verify the performance of the proposed EPPO in the single-user scenario. Note that the single-user scenario serves as a foundational analysis that allows us to isolate and analyze the core dynamics of the proposed method in a controlled environment. 

    \subsubsection{Comparison Results}
    \par Fig.~\ref{fig:simulation results of EPPO}\subref{fig:Cumulative rewards training curve.} illustrates the cumulative rewards for each episode of EPPO in comparison to other benchmark algorithms. In the initial stage, the rewards for all algorithms exhibit instability due to the poor performance of the UAV. This instability may arise from situations where the UAV flies outside the boundaries or reaches positions where it cannot establish LoS links. Subsequently, as the algorithms approach convergence, the rewards progressively rise and stabilize, indicating that an optimal strategy has been found.
    
    \par Moreover, the reward of the EPPO algorithm experiences substantial improvement during the mid-training phase. This phenomenon can be attributed to the energy consumption of the UAV, which will be explained in the following part. Additionally, this figure clearly shows that the EPPO algorithm achieves higher rewards, faster convergence, and improved stability compared to other algorithms. This can be explained by several factors. First, the IRS phase shift control strategy enhances the communication rate. Second, the integration of mogrifier LSTM strengthens the temporal information, thereby accelerating algorithm convergence. Finally, the inclusion of NECSA reduces the likelihood of the UAV entering prohibited areas. In summary, the EPPO algorithm improves data rates and mitigates excessive energy consumption.
    
    \par We plot the average rate for the EPPO algorithm and other benchmark algorithms in Fig.~\ref{fig:simulation results of EPPO}\subref{fig:Average rate training curve.} to illustrate the communication performance.. As can be seen, the proposed EPPO algorithm consistently maintains higher average data rates compared to other benchmark algorithms. The other algorithms initially exhibit low average rates, while as the algorithms converge, the average rates gradually increase to reach a stable value, which is still lower than that of EPPO. This can be explained by the fact that while the benchmark algorithms need to learn IRS phase shifts, EPPO does not require this phase shift learning process. The IRS phase control strategy computes phase shifts with greater precision.
    
    \par Furthermore, we evaluate the performance of these algorithms in solving the formulated problem with the energy consumption of the UAV in Fig.~\ref{fig:simulation results of EPPO}\subref{fig:Cumulative energy consumption training curve.}. As shown, during the training process of all algorithms, the energy consumption of the UAV converges to a stable state with low energy consumption. However, EPPO goes further by learning an even lower energy consumption strategy. One possible reason is that EPPO uses the IRS phase shift control strategy to quickly and accurately compute the IRS phase shift, which allows the action space to contain only the flight distance of the UAV, thereby enabling the agent to more efficiently find rational flight trajectories. According to Eq.~(\ref{eq:reward function}), this also accounts for the improvement of the EPPO reward observed in Fig.~\ref{fig:simulation results of EPPO}\subref{fig:Cumulative rewards training curve.}.
    

    \begin{figure*}[htbp]
    \centerline{\includegraphics[width=7in]{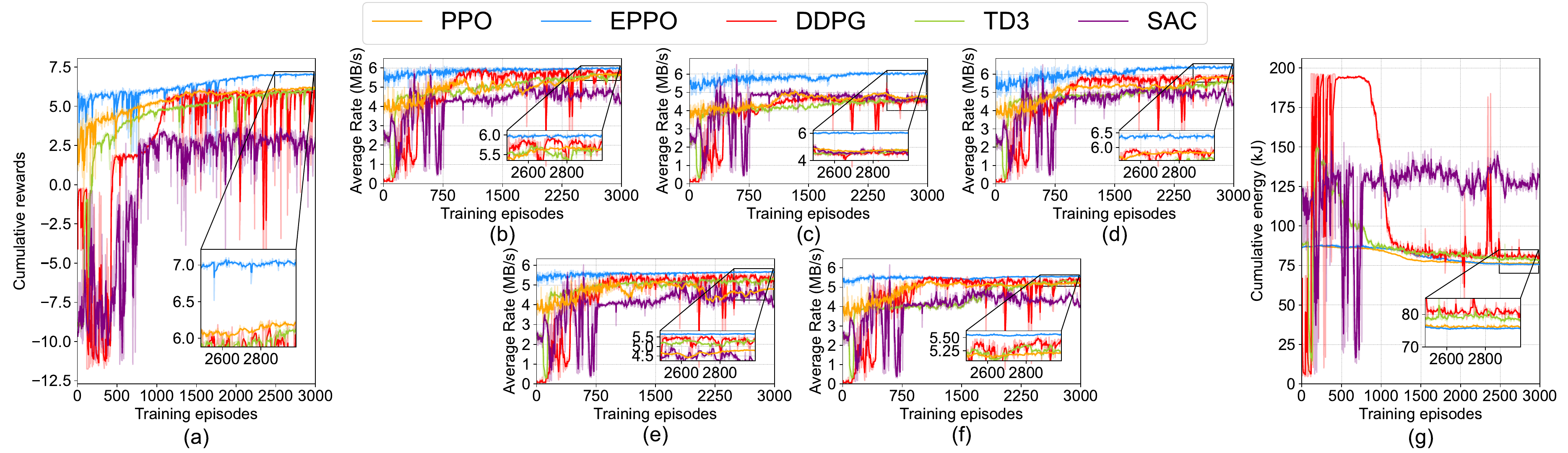}}
    		\caption{Muti-user training result. (a) Cumulative reward training curve. (b) Average rate training curve for the first user. (c) Average rate training curve for the second user. (d) Average rate training curve for the third user. (e) Average rate training curve for the fourth user. (f) Average rate training curve for the fifth user. (g) Cumulative energy training curve.}
    		\label{fig:Muti-user training result.}
    \end{figure*}

    \begin{figure}
    		\centerline{\includegraphics[width=3.5in]{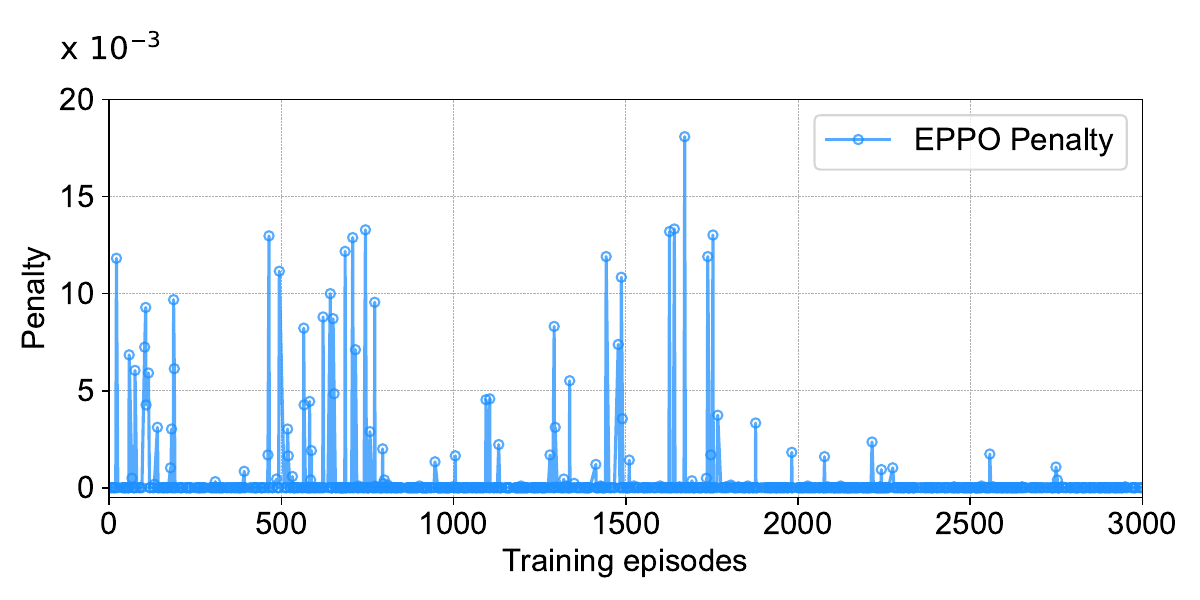}}
    		\caption{Average penalty curve in muti-user scenario.}
    		\label{fig:Average penalty curve in muti-user scenario.}
    \end{figure}

  

    \subsubsection{Ablation Simulation Results}
    \par Following this, we present comprehensive simulation results that illustrate the performance enhancements achieved through the integration of each individual improvement on the original algorithm. To verify the effectiveness of mogrifier LSTM, the conventional LSTM and GRU are introduced in this simulation. Fig.~\ref{fig:comparsion results of improvement}\subref{fig:Comparison of mogrifier LSTM, LSTM and GRU.} illustrates the rewards when PPO is combined with different LSTM networks. Mogrifier LSTM provides slight improvements in both convergence speed and final rewards. PPO with LSTM and PPO with GRU, on the other hand, exhibit similar performance to the PPO algorithm, without any notable improvement. This can be attributed to the unique architecture of mogrifier LSTM, which facilitates initial interactions between inputs and states, thereby enhancing its contextual modeling capabilities.
    
    \par Fig.~\ref{fig:comparsion results of improvement}\subref{fig:Comparison of PPO and PPO with NECSA.} shows the comparison between PPO and PPO with NECSA. Specifically, PPO with NECSA converges around the 750th episode, whereas PPO requires more time to converge. It is evident from the figure that the addition of NECSA results in an accelerated convergence speed for PPO. This is because NECSA maps continuous states to higher-level abstract states, enabling the agent to learn and generalize more efficiently. Moreover, Fig.~\ref{fig:comparsion results of improvement}\subref{fig:Comparison of PPO and PPO with phase control.} compares the capacity of PPO and PPO with IRS phase shift control. Clearly, PPO with IRS phase shift control shows a significant improvement in rewards. This result demonstrates the effectiveness of the IRS phase shift control strategy, which contributes to enhancing the communication data rate. Additionally, both IRS phase shift control and mogrifier LSTM exhibit favorable effects not only when used individually but also when combined, resulting in unexpected improvements. As shown in Fig.~\ref{fig:comparsion results of improvement}\subref{fig:PPO_PPO_PhaseControl_MogrifierLSTM.}, after the joint utilization of IRS phase shift control strategy and mogrifier LSTM, the algorithm experiences an improvement in rewards during the later stages of training, as explained in Fig.~\ref{fig:simulation results of EPPO}\subref{fig:Cumulative energy consumption training curve.} earlier.
    
    \subsubsection{Trajectory Visualization Results}
    
    \par Finally, we present the 3D trajectory plot obtained by the EPPO algorithm for the UAV. Fig.~\ref{fig:UAV 3D Trajectory of EPPO for single user.} shows the trajectory of UAV-carried IRS in single-user scenario. As can be seen, the UAV carries an IRS that initiates its service to the user from the upper-left corner, flying towards optimal positions to establish LoS links with both the user and the SU, thereby enhancing communication performance. Once the UAV identifies the appropriate location, it gradually approaches the user, as discussed in~\cite{Wu2021tutorial}, which is a reasonable approach.

\subsection{Optimization Results in Multi-User Scenario}
    \label{subsec:Optimization Results for muti-user}

     \begin{figure*}
    		\centerline{\includegraphics[width=7in]{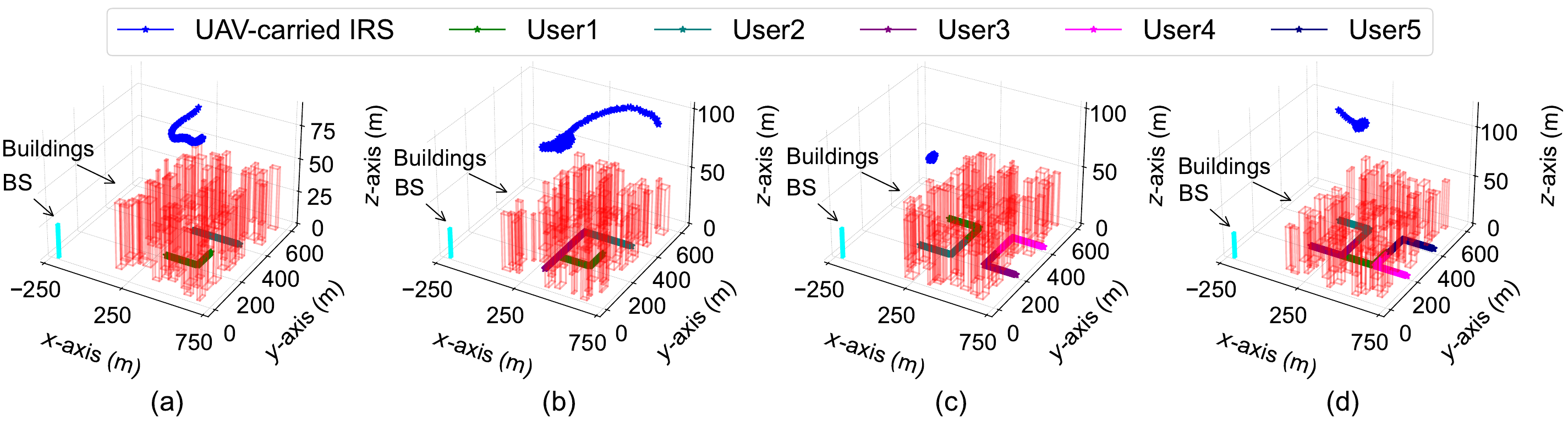}}
    		\caption{3D trajectories of the UAV under different numbers of users. (a) Trajectory of UAV for 2 users. (b) Trajectory of UAV for 3 users. (c) Trajectory of UAV for 4 users. (d) Trajectory of UAV for 5 users.}
    		\label{fig:UAV 3D Trajectory of EPPO different user number.}
    \end{figure*}


    \par In this subsection, the multi-user scenario is simulated to evaluate the performance of the proposed framework under more realistic and challenging conditions. Unlike the single-user scenario, the multi-user scenario introduces additional complexity due to the movements of the users and the need for efficient resource allocation among multiple competing users. Note that this case incorporates the Jain Fairness Index.

    \subsubsection{Comparison Results}
    
    \par Fig.~\ref{fig:Muti-user training result.}(a) offers a clear comparison of the five algorithms previously mentioned with respect to cumulative reward. As seen in the figure, during the early training stages, the proposed EPPO algorithm experiences slight fluctuations while maintaining a higher reward level, while the conventional PPO algorithm performs slightly worse than EPPO. In contrast, SAC, TD3, and DDPG exhibit significant fluctuations at a lower reward level. As the iterations proceed, the performance of EPPO significantly outperforms other algorithms. This is because the introduced NECSA mechanism significantly reduces the probability of the UAV encountering out-of-bounds occurrences and, consequently, reduces the out-of-bounds penalties as shown in Fig.~\ref{fig:Average penalty curve in muti-user scenario.}, thereby raising the lower limit of rewards in the early stages of training. Subsequently, the Jain Fairness Index indirectly guides the UAV to locations with fairness, which allows the IRS to better serve users and enhance cumulative rewards.

    \par Additionally, to validate the effectiveness of EPPO, we also plot the training curves for the data rate of each user. Specifically, Fig.~\ref{fig:Muti-user training result.}(b) illustrates the data rate for the first user, with the EPPO algorithm achieving slightly higher rates than PPO and outperforming other algorithms. In Figs.~\ref{fig:Muti-user training result.}(c),~\ref{fig:Muti-user training result.}(d),~\ref{fig:Muti-user training result.}(e), and~\ref{fig:Muti-user training result.}(f), the data rates of other users under different algorithms are depicted, which show a trend similar to that of the first user. EPPO exhibits better performance in convergence speed, data rate, and stability compared to PPO. This is attributed to the Jain Fairness Index, which guides the UAV towards a region with similar benefits but with a favor for fairness, thereby enhancing the data rate for the third user and strengthening communication stability for this user.

    \par Fig.~\ref{fig:Muti-user training result.}(g) presents the performance of various algorithms in terms of UAV energy consumption. The SAC algorithm stabilizes at a relatively high energy consumption with continuous training, while DDPG and TD3 show an initial increase in energy consumption, followed by stabilization in a low range, and PPO and EPPO exhibit smooth and stable convergence to a low value. It is noteworthy that EPPO follows a similar trend to PPO but converges to an even lower level of energy consumption. This is because the Jain Fairness Index encourages the system to intelligently utilize resources, guiding the UAV to locations with similar communication benefits but better fairness. This reduces the probability of the UAV flying to places where LoS links cannot be established, thereby reducing the UAV energy consumption.
    
    \subsubsection{Trajectory Visualization Results}
    
    \par Figs.~\ref{fig:UAV 3D Trajectory of EPPO different user number.}(a),~\ref{fig:UAV 3D Trajectory of EPPO different user number.}(b),~\ref{fig:UAV 3D Trajectory of EPPO different user number.}(c), and~\ref{fig:UAV 3D Trajectory of EPPO different user number.}(d) illustrate the trajectory of the UAV in multi-user scenarios. Initially, the UAV is randomly generated over the city. As time progresses, the UAV gradually moves closer to and hovers above multiple users from an edge position. There are two reasons for the continuous hovering of the UAV during this process. First, hovering at a certain speed consumes less energy than staying stationary in the air. Second, the UAV constantly adjusts its position to ensure fairness for every user.

    \begin{figure}
        \centerline{\includegraphics[width=3.5in]{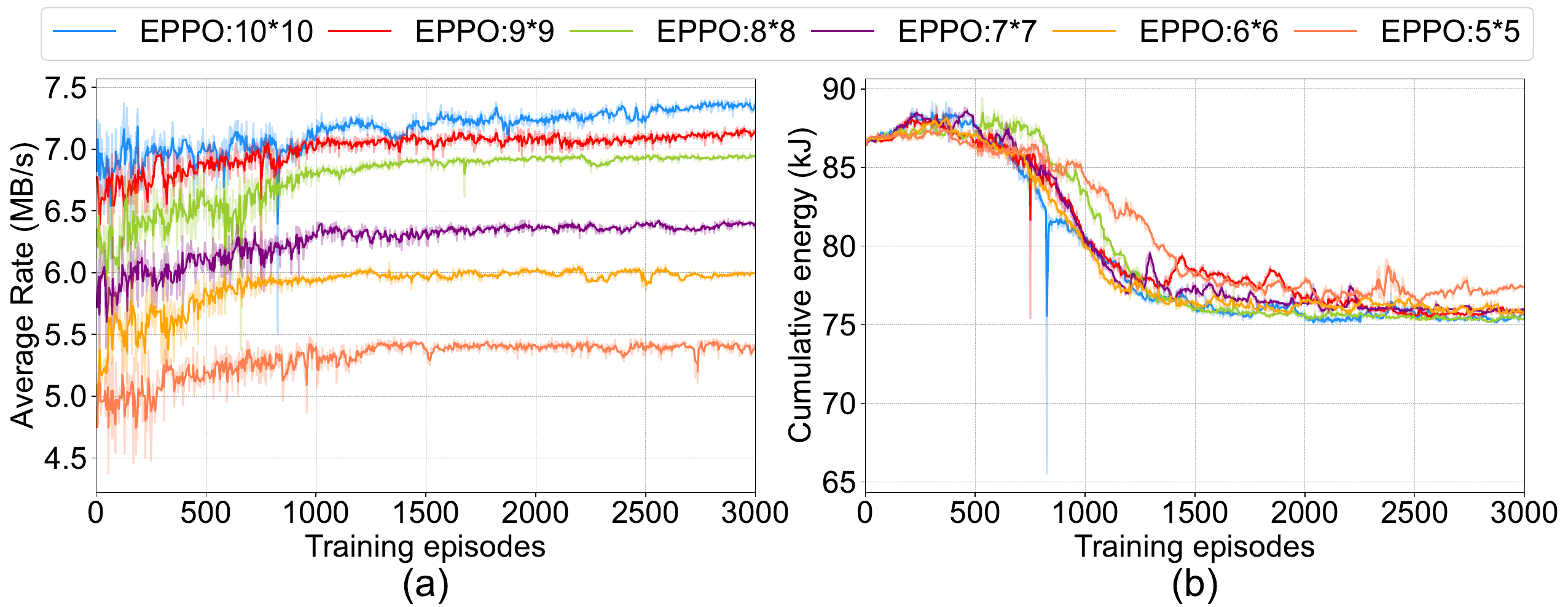}}
        \caption{The impacts for different IRS reflection element numbers. (a) Average rate curve of different IRS reflection element numbers. (b) Cumulative Energy curve of different IRS reflection element numbers.}
        \label{fig:The impacts for different IRS reflection element numbers.}
    \end{figure}
    \begin{figure}
        \centerline{\includegraphics[width=3.5in]{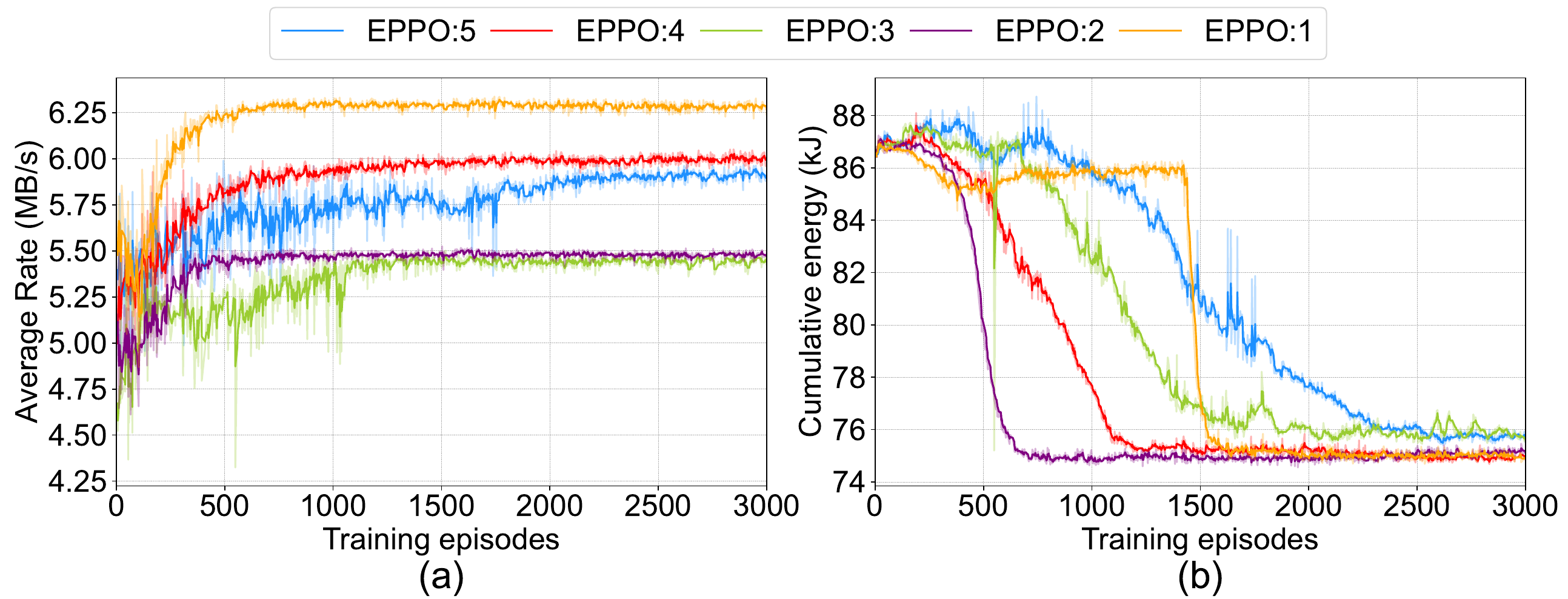}}
        \caption{The impacts for different user numbers. (a) Average rate curve of different user number. (b) Cumulative energy curve of different user numbers.}
        \label{fig:The impacts for different user numbers.}
    \end{figure}
    


    \subsubsection{Impacts of Simulation Parameters}
    
    \par Figs.~\ref{fig:The impacts for different IRS reflection element numbers.}(a) and~\ref{fig:The impacts for different IRS reflection element numbers.}(b) illustrate how varying the number of IRS reflection elements affects the performance of the EPPO algorithm. Specifically, it is observed that with an increase in the number of reflection elements, the data rates rise to different extents. This is because more reflection elements allow the system to finely adjust the direction and intensity of incident signals, thereby more effectively focusing the signals onto user devices. With an increase in the number of reflecting elements, the convergence speed for UAV energy consumption in the EPPO algorithm improves slightly, while energy consumption also decreases slightly.

    \par Figs.~\ref{fig:The impacts for different user numbers.}(a) and~\ref{fig:The impacts for different user numbers.}(b) show the impact of different numbers of users on the EPPO algorithm. It can be observed that the average rate of the system changes as the number of users varies. In our considered scenario, the base station communicates with users in a predefined order. In this case, the differences in average rate are likely influenced by the spatial distribution of the users and the optimized UAV trajectory. Specifically, as the number of users increases, certain users in dense urban low altitude communication scenarios may be blocked by buildings. At the same time, the optimized position of the UAV-carried IRS may fail to provide a LoS link to these blocked users, resulting in fluctuations in the average rate. Moreover, the number of users has a relatively small effect on the energy consumption of the UAV. As well known, there exists a speed at which the energy consumption is minimized. Fig.~\ref{fig:The impacts for different user numbers.}(b) shows that the UAV has learned to fly around speed at which the energy consumption is the lowest.

    \par In summary, EPPO consistently exceeds the performance of other algorithms regarding cumulative rewards, data rates, and energy efficiency in multi-user scenarios, while the Jain Fairness Index improves fairness and resource utilization.
    %
    %

    \section{Conclusion}
    \label{sec:Conclusion}
    
    \par In this paper, the UAV-carried IRS mmWave communication system has been investigated to address the challenges posed by obstacles blocking direct links between the SU and users in urban low altitude communication scenarios. By leveraging a UAV to carry an IRS, we effectively rebuilt communication links in such scenarios. Specifically, we formulated a joint optimization problem aimed at maximizing data rate and minimizing UAV energy consumption, while the problem is characterized by its high real-time and dynamic complexity. To tackle this, we proposed an EPPO algorithm that incorporating several enhancements such as neural episodic control, improved LSTM, and IRS phase shift control to boost stability and convergence speed. Simulation results confirmed the effectiveness of the EPPO algorithm, showing better performance than other benchmarks, especially in enhancing communication rates and reducing energy consumption. 




\bibliographystyle{IEEEtran}
\bibliography{main}

\vspace{-10mm}

\begin{IEEEbiography}[{\includegraphics[width=1in,height=1.25in,clip,keepaspectratio]{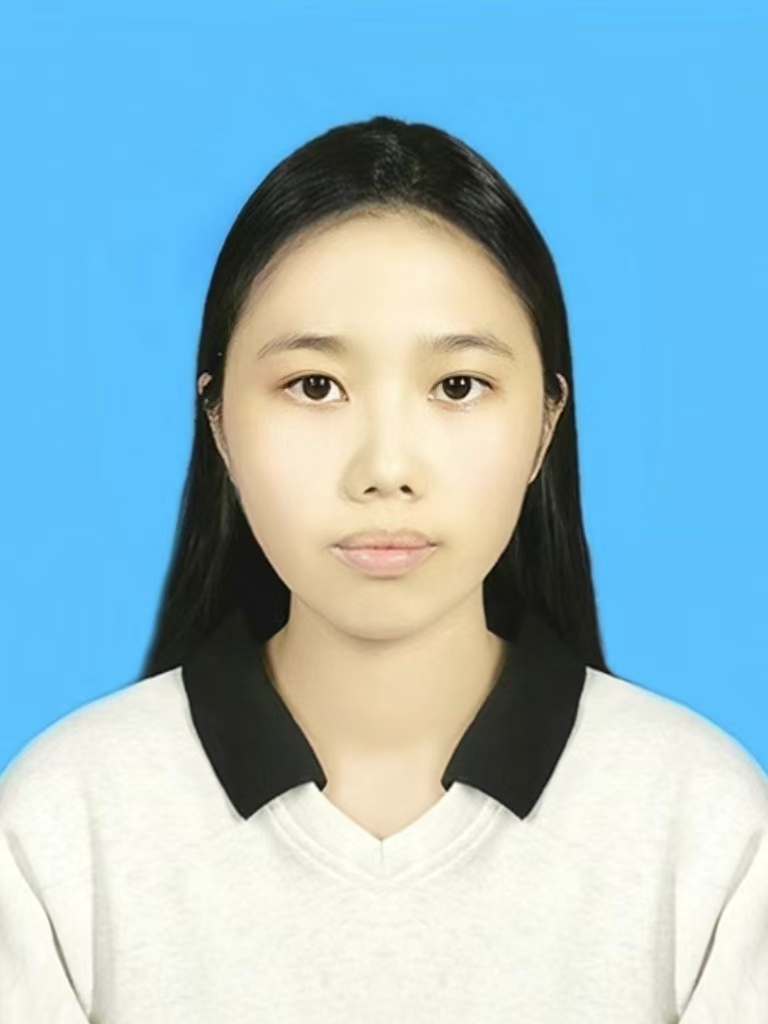}}]{Wenwen Xie} received the B.S. degree in Computer Science and Technology from Hefei University of Technology, Hefei, China, in 2022. She is currently working toward the MS degree in Computer Science and Technology at Jilin University, Changchun, China. Her research interests include UAV communications, IRS beamforming and deep reinforcement learning.

\end{IEEEbiography}

\vspace{-10mm}

\begin{IEEEbiography}[{\includegraphics[width=1in,height=1.25in,clip,keepaspectratio]{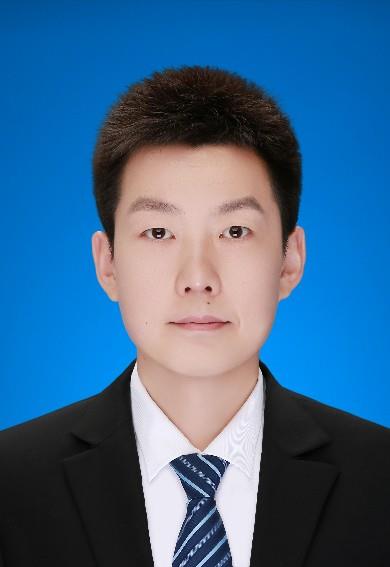}}]{Geng Sun} (S’17-M’19-SM’24) received the
B.S. degree in communication engineering from Dalian Polytechnic University, and the Ph.D. degree in computer science and technology from Jilin University, in 2011 and 2018, respectively. He was a Visiting Researcher with the School of Electrical and Computer Engineering, Georgia Institute of Technology, USA. He is a Professor in College of Computer Science and Technology at Jilin University, and his research interests include wireless networks, UAV communications, collaborative beamforming and optimizations.
\end{IEEEbiography}

\vspace{-10mm}

\begin{IEEEbiography}[{\includegraphics[width=1in,height=1.25in,clip,keepaspectratio]{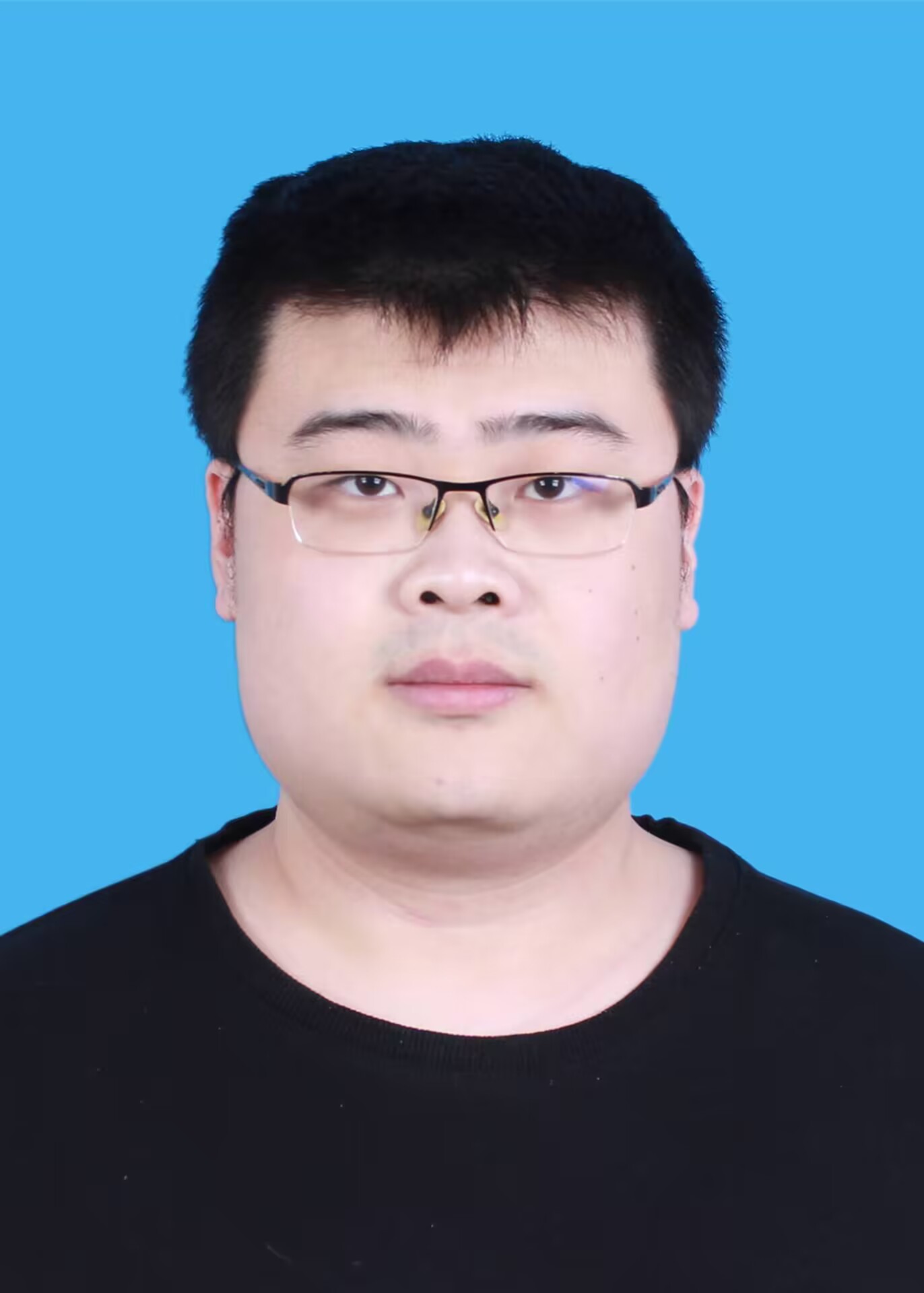}}]{Bei Liu}
received the B.S. degree in software engineering from Jilin University in 2021. He is currently working towards the M.S. degree at the College of Computer Science and Technology, Jilin University. His research interests include intelligent reflecting surface and deep reinforcement learning.
\end{IEEEbiography}

\vspace{-10mm}
\begin{IEEEbiography}
[{\includegraphics[width=1in,height=1.25in,clip,keepaspectratio]{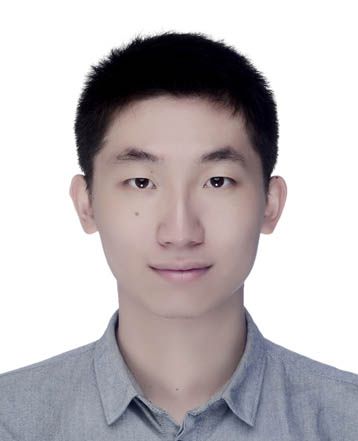}}]
{Jiahui Li} received his B.S. in Software Engineering, and M.S. and Ph.D. in Computer Science and Technology from Jilin University, Changchun, China, in 2018, 2021, and 2024, respectively. He was a visiting PhD student at the Singapore University of Technology and Design (SUTD). He currently serves as an assistant researcher in the College of Computer Science and Technology at Jilin University. His current research focuses on integrated air-ground networks, UAV networks, wireless energy transfer, and optimization.
\end{IEEEbiography}



\vspace{-10mm}

\begin{IEEEbiography}[{\includegraphics[width=1in,height=1.25in,clip,keepaspectratio]{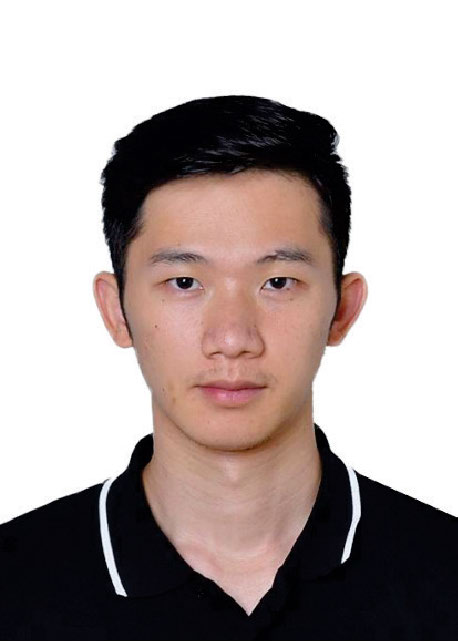}}]{Jiacheng Wang} received the Ph.D. degree from the School of Communication and Information Engineering, Chongqing University of Posts and Telecommunications, Chongqing, China. He is currently a Research Associate in computer science and engineering with Nanyang Technological University, Singapore. His research interests include wireless sensing, semantic communications, and metaverse.
\end{IEEEbiography}
\vspace{-10mm}
\begin{IEEEbiography}
[{\includegraphics[width=1in,height=1.25in,clip,keepaspectratio]{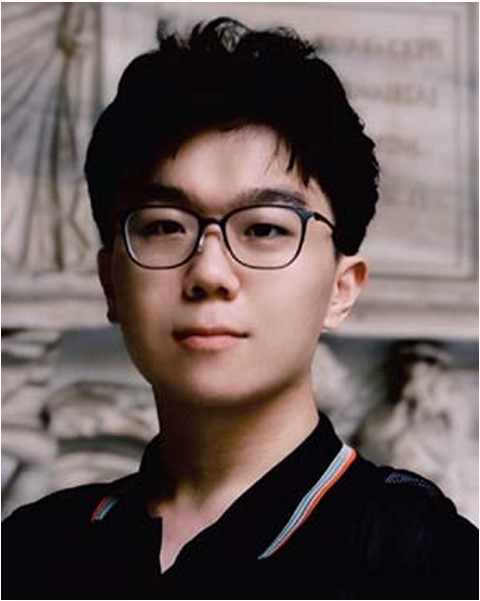}}]{Hongyang Du} (Graduate Student Member, IEEE) received the BSc degree from Beijing Jiaotong University, Beijing, China, in 2021. He is currently working toward the PhD degree with the School of Computer Science and Engineering, Energy Research Institute @ NTU, Nanyang Technological University, Singapore, under the Interdisciplinary Graduate Program. He is the Editor-in-chief assistant of IEEE Communications Surveys \& Tutorials (2022-2024). He was recognized as an exemplary reviewer of the IEEE Transactions on Communications and IEEE Communications Letters in 2021. He was the recipient of the IEEE Daniel E. Noble Fellowship Award from the IEEE Vehicular Technology Society, in 2022, the recipient of the IEEE Signal Processing Society Scholarship from the IEEE Signal Processing Society, in 2023, the recipient of Chinese Government Award for Outstanding Students Abroad, in 2023, and the recipient of the Singapore Data Science Consortium (SDSC) Dissertation Research Fellowship in 2023. He won the Honorary Mention award in the ComSoc Student Competition from IEEE Communications Society, in 2023, and the First and Second Prizes in the 2024 ComSoc Social Network Technical Committee (SNTC) Student Competition. His research interests include semantic communications, generative AI, and resource allocation.
\end{IEEEbiography}
\vspace{-10mm}
\begin{IEEEbiography}[{\includegraphics[width=1in,height=1.25in,clip,keepaspectratio]{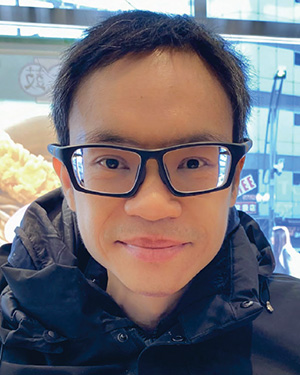}}]{Dusit Niyato} (Fellow, IEEE) received the B.Eng. degree from the King Mongkuts Institute of Technology Ladkrabang (KMITL), Thailand, in 1999, and the Ph.D. degree in electrical and computer engineering from the University of Manitoba, Canada, in 2008. He is currently a Professor with the School of Computer Science and Engineering, Nanyang Technological University, Singapore. His research interests include the Internet of Things (IoT), machine learning, and incentive mechanism design. 
\end{IEEEbiography}
\vspace{-10mm}
\begin{IEEEbiography}[{\includegraphics[width=1in,height=1.25in,clip,keepaspectratio]{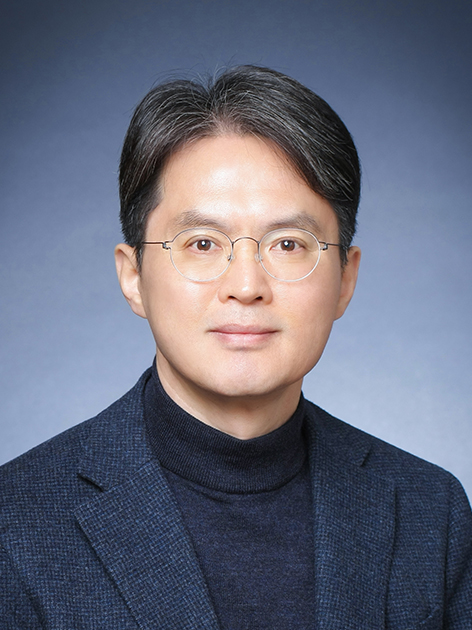}}]{Dong In Kim} (Fellow, IEEE) received the Ph.D. degree in electrical engineering from the University of Southern California, Los Angeles, CA, USA, in 1990. He was a Tenured Professor with the School of Engineering Science, Simon Fraser University, Burnaby, BC, Canada. He is currently a Distinguished Professor with the College of Information and Communication Engineering, Sungkyunkwan University, Suwon, South Korea. He is a Fellow of the Korean Academy of Science and Technology and a Member of the National Academy of Engineering of Korea. He was the first recipient of the NRF of Korea Engineering Research Center in Wireless Communications for RF Energy Harvesting from 2014 to 2021. He received several research awards, including the 2023 IEEE ComSoc Best Survey Paper Award and the 2022 IEEE Best Land Transportation Paper Award. He was selected the 2019 recipient of the IEEE ComSoc Joseph LoCicero Award for Exemplary Service to Publications. He was the General Chair of the IEEE ICC 2022, Seoul. Since 2001, he has been serving as an Editor, an Editor at Large, and an Area Editor of Wireless Communications I for IEEE Transactions on Communications. From 2002 to 2011, he served as an Editor and a Founding Area Editor of Cross-Layer Design and Optimization for IEEE Transactions on Wireless Communications. From 2008 to 2011, he served as the Co-Editor- in-Chief for the IEEE/KICS Journal of Communications and Networks. He served as the Founding Editor-in-Chief for the IEEE Wireless Communications Letters from 2012 to 2015. He has been listed as a 2020/2022 Highly Cited Researcher by Clarivate Analytics.
\end{IEEEbiography}
\end{document}